\documentclass[12pt]{article}

\pdfoutput=1

\usepackage{fullpage}
\usepackage{amsmath,amsfonts,amssymb,graphics,graphicx,fancyhdr,subfigure,float,fancybox,url,xr}
\usepackage{wrapfig}
\usepackage[T1]{fontenc}
\usepackage{cite}

\usepackage{rotating}
\usepackage{colortbl}
\usepackage{sidecap}
\usepackage{color} 
\usepackage[linktoc=all]{hyperref}
\providecommand\phantomsection{}

\makeatletter \renewcommand\@biblabel[1]{#1.} \makeatother
\newcommand{\e}{{\rm E}}

\newcommand{\bh}{\widehat}

\newcommand{\real}{\mathbb{R}}
\newcommand{\iid}{\stackrel{i.i.d}{\sim}}

\newcommand{ \bm }[1]{ \mbox{\bf {#1}} }
\newcommand{ \bmm }[1]{ \mbox{\boldmath ${#1}$} }
\newcommand{\bb}{\bm{b}}

\newcommand{\fdr}{{\rm FDR}}

\newcommand{\bw}{\bm{w}}

\newcommand{\bY}{\bm{Y}}

\newcommand{\bB}{\bm{B}}

\newcommand{\bD}{\bm{D}}
\newcommand{\bR}{\bm{R}}

\newcommand{\bE}{\bm{E}}

\newcommand{\bW}{\bm{W}}

\newcommand{\bV}{\bm{V}}

\newcommand{\bU}{\bm{U}}
\newcommand{\bC}{\bm{C}}

\newcommand{\by}{\bm{y}}

\newcommand{\bv}{\bm{v}}

\newcommand{\bzero}{\bm{0}}

\newcommand{\bGamma}{\bmm{\Gamma}}

\newcommand{\bgamma}{\bmm{\gamma}}

\newcommand{\no}{\noindent}
\newcommand{\bS}{\bm{L}}
\newcommand{\bL}{\bmm{z}}

\begin{document}

\title{Statistical Significance of Variables Driving Systematic Variation}
\author{Neo Christopher Chung\thanks{Lewis-Sigler Institute for Integrative Genomics, Princeton University, Princeton, NJ 08544, USA.} \ and John D. Storey\thanks{Lewis-Sigler Institute for Integrative Genomics \& Department of Molecular Biology, Princeton University, Princeton, NJ 08544, USA.} \thanks{To whom correspondence should be sent: John Storey, \url{http://genomine.org/contact.html}.}}
\date{July 2013}

\maketitle

\tableofcontents

\clearpage
\begin{abstract} 
There are a number of well-established methods such as principal components analysis (PCA) for automatically capturing systematic variation due to latent variables in large-scale genomic data.  PCA and related methods may directly provide a quantitative characterization of a complex biological variable that is otherwise difficult to precisely define or model.  An unsolved problem in this context is how to systematically identify the genomic variables that are drivers of systematic variation captured by PCA.  Principal components (and other estimates of systematic variation) are directly constructed from the genomic variables themselves, making measures of statistical significance artificially inflated when using conventional methods due to over-fitting.
We introduce a new approach called the {\em jackstraw} that allows one to accurately identify genomic variables that are statistically significantly associated with any subset or linear combination of principal components (PCs).  The proposed method can greatly simplify complex significance testing problems encountered in genomics and can be utilized to identify the genomic variables significantly associated with latent variables.  Using simulation, we demonstrate that our method attains accurate measures of statistical significance over a range of relevant scenarios. We consider yeast cell-cycle gene expression data, and show that the proposed method can be used to straightforwardly identify genes that are cell-cycle regulated with an accurate measure of statistical significance. We also analyze gene expression data from post-trauma patients, allowing the gene expression data to provide a molecularly-driven phenotype.  Using our method, we find a greater enrichment for inflammatory-related gene sets compared to the original analysis that utilizes a clinically defined, although likely imprecise, phenotype. The proposed method provides a useful bridge between large-scale quantifications of systematic variation and gene-level significance analyses.
\end{abstract}

\ 

\noindent Abbreviations: KS test, Kolmogorov-Smirnov test; PCA, principal components analysis; PC, principal component

\clearpage
\section*{Introduction} \addcontentsline{toc}{section}{Introduction}
Latent variable models play an important role in understanding variation in genomic data \cite{Price2006,leek2007chg}.  They are particularly useful for characterizing systematic variation in genomic data whose variable representation is unobserved or imprecisely known (Fig. \ref{MolecularModel}).   Principal component analysis (PCA) has proven to be an especially informative method for capturing quantitative signatures of latent variables in genomic data, and it is in widespread use across a range of applications.  For example, PCA has been successfully applied to uncover the systematic variation in gene expression \cite{Raychaudhuri2000, Holter2000, Alter2000}, estimate structure in population genetics \cite{Zhu2002, Price2006}, and account for dependence in multiple hypothesis testing \cite{leek2007chg, leek2008gfm}. Generally, principal components (PCs) can be thought of as estimates of unobserved manifestation of latent variables; they are constructed by aggregating variation across thousands or more genomic variables \cite{jolliffe2002principal}. What is missing from this highly successful system is a method to precisely identify which genomic variables are the statistically significant drivers of the PCs in genomic data, which in turn identifies the genomic variables associated with the unobserved latent variables.  

In a typical application of PCA, all genomic variables (e.g. microarray probes, genetic loci, etc.) will have nonzero ``loadings'', meaning that they all make some contribution to the construction of PCs.  In some cases, when many (or most) of these contributions are forcibly set to zero, similar PCs nevertheless emerge.  Methods have been proposed to induce sparsity in the loadings\footnote{The difference between our goal and the goal of sparsity is analogous to the difference between penalized regression and multiple hypothesis testing, both of which have been proven to be useful in genomics and other applications \cite{ESL2011}.}, for example, with a lasso penalized PCA or a Bayesian prior  \cite{Jolliffe2003, Zou2006, Witten2009, Engelhardt2010}.  Also, various formulations of statistical significance have been considered previously in the context of PCA.  These have usually been focused on scenarios where the number of observations is substantially larger than the number of variables, significance is measured in terms of a completely unstructured data matrix where all variables are mutually independent, or the goal is to only determine the number of significant principal components \cite{Girshick:1939p2599, Anderson1963, Buja1992, Tracy1996, Johnstone2001, PeresNeto:2003p2597, Timmerman:2010p2598, Linting:2011p2571}.  The problem we consider here differs from those scenarios.

Our goal is not a minimal representation of a PCA; we would like instead to develop a strategy that accurately identifies which genomic variables are truly associated with systematic variation of interest.  This can be phrased in statistical terminology as developing a significance test for associations between genomic variables and a given set, subset, or linear combination of PCs estimated from genomic data.   We introduce a new resampling approach, which we call the \emph{jackstraw}, to rigorously identify the genomic variables associated with PCs of interest, as well as subsets and rotations of PCs of interest. Our approach is capable of obtaining the empirical null distribution of association statistics (e.g., F-statistics) and applying these to the observed association statistics between genomic features and PCs in order to obtain valid statistical significance measures. Succinctly, new PCs are computed from a dataset with a few independently permuted variables, which become tractable ``synthetic'' null variables.  The association statistics between newly computed PCs and synthetic null variables serve as empirical null statistics, accounting for the measurement error and over-fitting of PCA. 

As an application, we consider the problem of identifying genes whose expression is cell-cycle regulated.  In this case, there are infinitely many theoretical curves that would represent ``cell-cycle regulation'' to the point where a standard statistical analysis involves an unwieldy ``composite null hypothesis'' \cite{lehmann1997testing}.  We identify the few realized patterns of cell-cycle regulated gene expression through PCA and we are able to directly test whether each gene is associated with these using the proposed approach. As another application, we analyzed observational gene expression profiles of blunt-force trauma patients \cite{Desai2011}, whose post-trauma inflammatory responses are difficult to be quantified using conventional means. When the clinical phenotype of interest can not be precisely measured and modeled, we may estimate it directly from genomic data itself. We identify genes driving systematic variation in gene expression of post-trauma patients and demonstrate that our analysis is biologically richer than the original analysis \cite{Desai2011}.

PCA has direct connections to Independent Component Analysis \cite{ESL2011} and $K$-means clustering \cite{Zha2001, Ding2004}.  Therefore, the methods we propose are likely applicable to those models as well.  Furthermore, this approach has potential generalizations to a much broader class of clustering and latent variable methods that all seek to capture systematic variation.

\section*{Statistical Model and Approach}  
\begin{figure*}[h]
\begin{center}
    \includegraphics[width=.8\textwidth,natwidth=2399,natheight=2099]{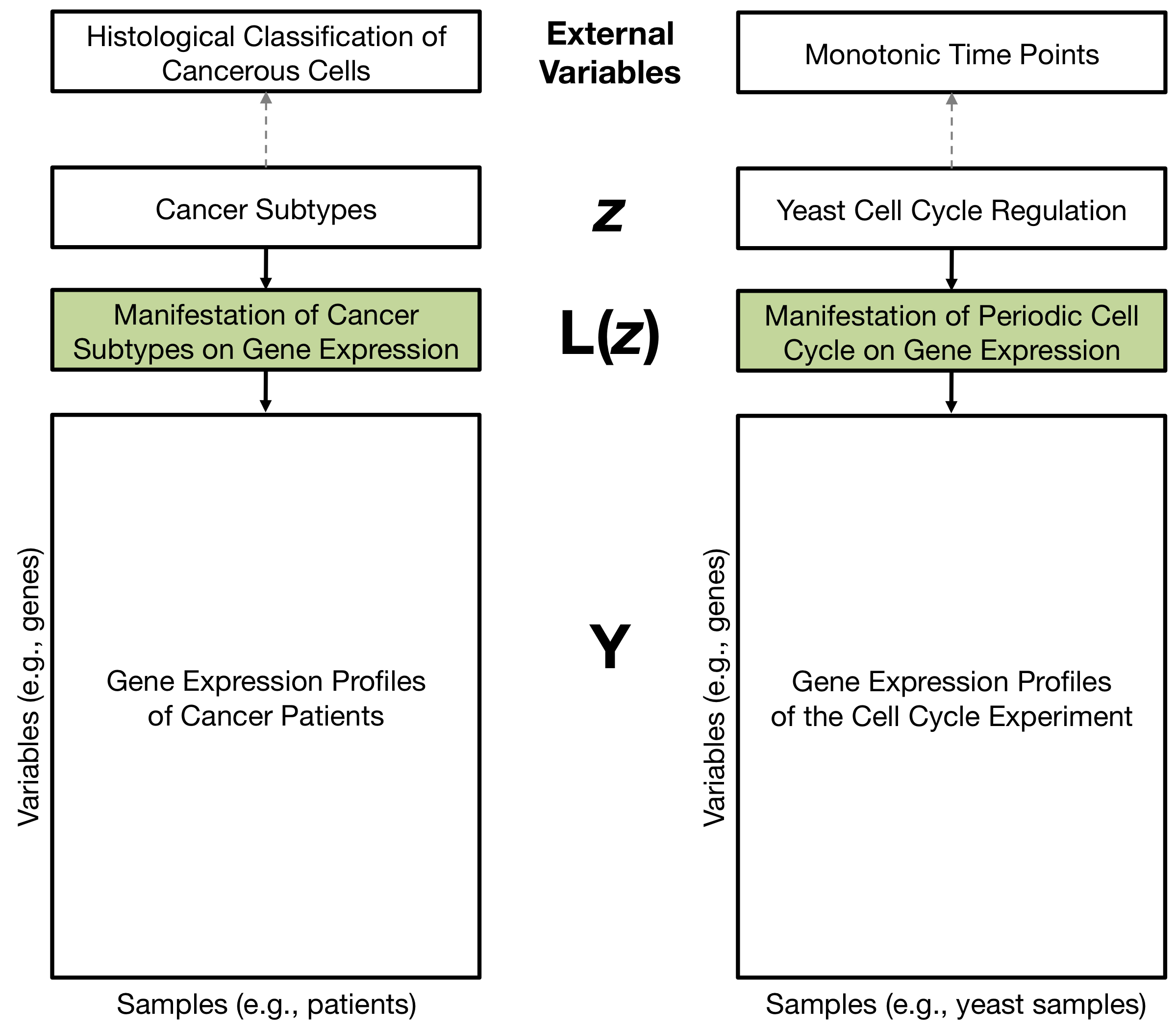}
	\caption{{\em Illustration of systematic variation genomic data due to latent variables. Complex biological variables, such as cancer subtypes and cell cycle regulation, may be difficult to define, measure, or model. Instead, we can characterize the manifestation of latent variables, $\bS(\bL)$, directly from high-dimensional genomic data using PCA and related methods. The proposed method calculates the statistical significance of associations between variables in $\bY$ and estimates of $\bS$, while accounting for over-fitting due to the fact that $\bS$ must be estimated from $\bY$.}}
\label{MolecularModel}
\end{center}
\end{figure*}

\addcontentsline{toc}{section}{Statistical Model and Approach}
Consider a $m \times n$ row-centered expression data matrix $\bY$ with $m$ observed variables measured over $n$ observations ($m \gg n$). $\bY$ may contain systematic variation across the variables from an arbitrarily complex function of latent variables $\bL$. We may calculate the expected influence of the latent variables on $\bY$ by $\e[\bY|\bL]$, and then write $\bY = \e[\bY|\bL] + \bE$, where $\bE$ is defined as $\bY - \e[\bY|\bL]$. There exists a $r \times n$ matrix, called $\bS(\bL)$, that is a row basis for $\e[\bY|\bL]$, where $r \leq n$ \cite{leek2007chg,leek2008gfm}. This low dimensional matrix $\bS(\bL)$ can be thought of as the manifestation of the latent variables in the genomic data.  As illustrated in Fig. \ref{MolecularModel}, this conditional factor model is common for biomedical and genomic data \cite{Leek2010}.  Since $\bL$ is never directly observed or utilized in the model, we will abbreviate  $\bS(\bL)$ as $\bS$.   This yields the model 
\begin{equation} \label{eq:mod1}
\bY = \bB\bS + \bE
\end{equation}
where $\bB$ is a $m \times r$ matrix of unknown parameters of interest.  The $i^\textnormal{th}$ row of $\bB$, which we write as $\bb_i$, quantifies the relationship between the latent variable basis $\bS$ and genomic variable $\by_i$. This model \eqref{eq:mod1} is schematized in Fig. \ref{LatentVariableModel}.

The principal components (PCs) of $\bY$ may be calculated by taking the singular value decomposition (SVD) of $\bY$.  This yields $\bY = \bU\bD\bV^T$ where $\bU$ is a $m \times n$ orthonormal matrix, $\bD$ is a $n \times n$ diagonal matrix, and $\bV$ is a $n \times n$ orthonormal matrix. PCs are then the rows of $\bD\bV^T$, where the $i^\textnormal{th}$ PC is found in the $i^\textnormal{th}$ row of $\bD\bV^T$.   The columns of $\bU$ are considered to be the loadings of their respective PCs.

Suppose that the row-space of $\bS$ has dimension $r$.  The top $r$ PCs may then be utilized to estimate the row basis for $\bS$ \cite{jolliffe2002principal}.  Specifically, under a mild set of assumptions, it has been shown that as $m \rightarrow \infty$, the top $r$ PCs of $\bY$ converge with probability $1$ to a matrix whose row space is equivalent to that of $\bS$ \cite{Leek2010}.  For our estimation purposes, we only need to consider the $\bV^T$ matrix since this captures the row-space.  We would therefore estimate $\bS$ by simply obtaining the top $r$ right singular vectors, which we denote by $\bV^T_r$.

Let's now consider a concrete example of $\bL$, $\bS$, $\bV^T_r$, and the ultimate inference goal.  \cite{Spellman1998} carried out a gene expression study to identify cell cycle-regulated genes of \emph{Saccharomyces cerevisiae} (Fig. \ref{AlterSVD}).  In this experiment, $m=5981$ genes' expression values were measured over $n=14$ time points in a culture of yeast cells whose cell cycles had been synchronized.  Here, $\bL$ are the latent variables that represent the dynamic gene expression regulatory program over the yeast cell cycle.  $\bS$ is the manifested influence of $\bL$ on the observed scale of gene expression measurements (Fig. \ref{MolecularModel}).  The ordered time points themselves do not capture the underlying cell cycle regulation, and it is therefore not clear how to {\em a priori} accurately model $\bS$.  If $\bS$ were directly observed, then we could identify which genes are cell-cycle regulated by performing a significance test of $H_0: \bb_i = \bzero$ vs. $H_1: \bb_i \not= \bzero$ for each gene $i$.

However, since $\bS$ is not observed, we can instead perform the analogous association test using $\bV^T_r$.  Fig. \ref{AlterSVD}(a) shows the first two PCs of $\bY$, where it can be seen that these capture systematic variation that resembles cell-cycle regulation.  (It should be noted that the remaining PCs, three and higher, do not capture any systematic variation of interest.)  Since the row-spaces of $\bS$ and $\bV^T_r$ ($r=2$) are theoretically close \cite{Leek2010}, we can instead utilize the model
\begin{equation}  \label{eq:mod2}
\bY = \bGamma\bV^T_r + \bE^{\prime},
\end{equation}
where $\bGamma$ is a $m \times r$ matrix of unknown coefficients. We would then perform a significance test of $H_0: \bgamma_i = \bzero$ vs. $H_1: \bgamma_i \neq \bzero$ for each gene $i$.

\begin{figure*}
\begin{center}
    \includegraphics[width=.8\textwidth]{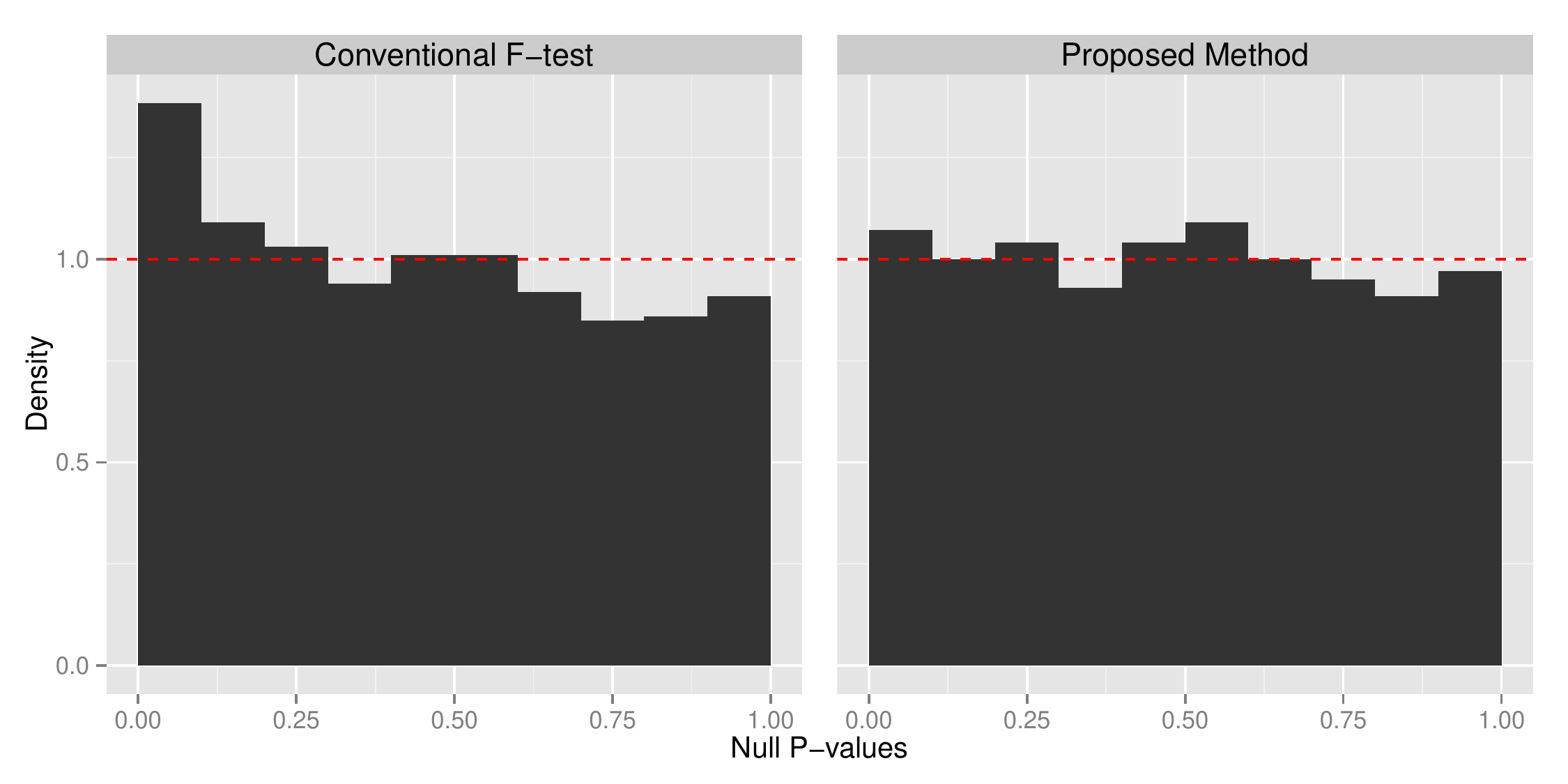}
	\caption{{\em Evaluation of significance measures of associations between variables and their PCs by comparing true null p-values and the Uniform(0,1) distribution. {\rm(a)} The conventional F-test results in anti-conservative p-values, as demonstrated by null p-values being skewed towards $0$. {\rm(b)} The proposed method produces null p-values distributed Uniform(0,1). A red dashed line shows the Uniform(0,1) density function.}}
\label{SimulationExample}
\end{center}
\end{figure*}

Note that if $\bV^T_r \rightarrow \bS$ in row-space as $m \rightarrow \infty$, then these two hypothesis tests would be equivalent.  However, for fixed $m$, they are not equivalent.  There are two main issues: (i) $\bV^T_r$ is a noisy estimate of $\bS$; (ii) $\bV^T_r$ is itself a function of $\bY$, so hypothesis testing on $\bY = \bGamma\bV^T_r + \bE^{\prime}$ results in over fitting as shown in the left panel of Fig. \ref{SimulationExample}.  Our proposed method deals with problem (ii) by accounting for the over-fitting that is intrinsic to performing hypothesis testing on model \eqref{eq:mod2}.  The numerical results in this paper are carried out so that we generate the data from model \eqref{eq:mod1} and evaluate the accuracy of the significance based on the truth from model \eqref{eq:mod1}.  Therefore, we provide direct evidence that our proposed method accounts for both issues (i) and (ii).  

\section*{Proposed Algorithms}   
\begin{figure*}
\begin{center}
    \includegraphics[width=.8\textwidth]{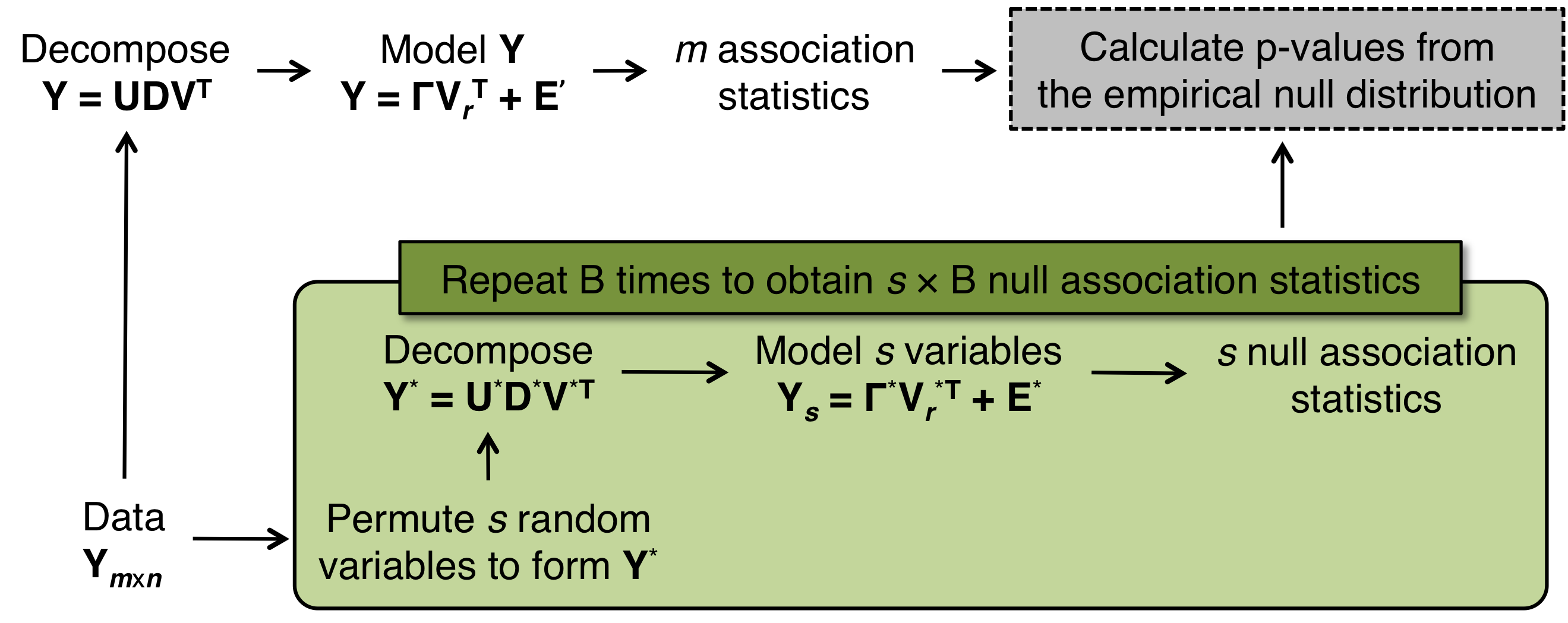}
	\caption{{\em A schematic of the general steps of the proposed algorithm to calculate the statistical significance of associations between variables (rows in $\bY$) and their top $r$ principal components ($\bm{V}^T_r$). By independently permuting a small number ($s$) of variables and recalculating the PCs, we generate tractable ``synthetic'' null variables while preserving the overall systematic variation. Association statistics between the $s$ synthetic null variables in $\bY^*$ and $\bm{V}^{*T}_r$ form the empirical null distribution, automatically taking account over-fitting intrinsic to testing for associations between a set of observed variables and their PCs.}}
\label{AlgorithmScheme}
\end{center}
\end{figure*}

\addcontentsline{toc}{section}{Proposed Algorithms}
We have developed a resampling method (Fig. \ref{AlgorithmScheme}) to obtain accurate statistical significance measures of the associations between observed variables and their PCs, accounting for the over-fitting characteristics due to computation of PCs from the same set of observed variables. The proposed algorithm replaces a small number $s$ ($s \ll m$) of observed variables with independently permuted ``synthetic'' null variables, while preserving the overall systematic variation in the data.  We denote the new matrix with the $s$ synthetic null variables replacing their original values as $\bY^*_{m \times n}$.  This is simply the original matrix $\bY$ with the $s$ rows of $\bY$ replaced by independently permuted versions.  On each permutation data set $\bY^*$, we calculate association statistics for each synthetic null variable, exactly as was done on the original data.  We carry this out $B$ times, effectively creating $B$ sets of permutation statistics.  The association statistics calculated on $\bY$ are then compared to the association statistics calculated on {\em only the s synthetic null rows of} $\bY^*$ to obtain statistical significance measures.  

\

\no \underline{Algorithm to Calculate Significance of Variables Associated with PCs}
\begin{enumerate}
\item[1.] Obtain $r$ PCs of interest, $\bV^T_r$ by applying SVD to the row-centered matrix $\bY_{m \times n} = \bU\bD\bV^T$.
\item[2.] Calculate $m$ observed F-statistics $F_1, \ldots ,F_m$, testing $H_0: \bgamma_i = \bzero$ vs. $H_1: \bgamma_i \neq \bzero$ from model \eqref{eq:mod2}.
\item[3.] Randomly select and permute $s$ rows of $\bY_{m \times n}$, resulting in $\bY^*_{m \times n}$.
\item[4.] Obtain $\bV^{*T}_r$ from SVD applied to $\bY^* = \bU^*\bD^*\bV^{*T}$.
\item[5.] Calculate null F-statistics $F^{0b}_1, \ldots, F^{0b}_s$ from the $s$ synthetic null rows of $\bY^*$ as in Step 2, where $\bV^T_r$ is replaced with $\bV^{*T}_r$.
\item[6.] Repeat steps 3-5 $b = 1, \ldots, B$ times to obtain a total $s \times B$ of null F-statistics.
\item[7.] Compute the p-value for variable $i$ ($i=1,\ldots,m$) by: $$p_i = \frac{\#\{F^{0b}_j \geq F_i; j=1, \ldots, s, b=1, \ldots, B\}}{s \times B}$$
\item[8.] Identify statistically significant tests based on the p-values $p_1, p_2, \ldots,$ $p_m$ (e.g., using false discovery rates).
\end{enumerate}

We call this approach the \emph{jackstraw} for the following reason.  By permuting a relatively small amount of observed variables in the original matrix, the underlying systematic variation due to latent variables is preserved as a whole. This makes the PCs of $\bY^*$ almost identical to the PCs of the original data, $\bY$, up to variation due to over-fitting of the noise.  Replacing $s$ variables with null versions is analogous to the game of jackstraws where the goal is to remove one stick at a time from a structured set of sticks without disrupting the overall structure of the sticks.  Since the overall structure of $\bY$ is preserved in $\bY^*$, we know that the level of associations between these synthetic null variables and the top $r$ PCs is purely due to the over-fitting nature of PCA.  From these synthetic null statistics, we can therefore capture and adjust for the over-fitting among the original statistics. 

A balance between the number of resampling iterations $B$ and the number of synthetic null variables $s$ is relevant to the speed of the algorithm and the accuracy of the resulting p-values.  One extreme is to set $s=1$, where the accuracy of the p-values is maximized while the algorithm is the least efficient. In each resampling iteration, $s$ determines the number of estimated null statistics, so to get the same resolution of a particular empirical null distribution ($s \times B$ total null statistics), $B$ must increase proportionally with a decreasing $s$.  For example, $s=1$ and $B=10000$ yields the same number of null statistics as $s=100$ and $B=100$.  The number of true null variables in $\bY^*$ is always greater than or equal to the number of true null variables in the original matrix $\bY$. Therefore an increase of $s$ in the proposed algorithm may lead to a greater over-fitting into the noise of $\bY^*$ relative to the over-fitting in $\bY$, resulting in conservative estimates of significance.  Due to this favorable trade-off between $s$ and $B$, the proposed algorithm is always guarded against anti-conservative bias.

The hypothesis test $H_0: \bgamma_i = \bzero$ vs. $H_1: \bgamma_i \neq \bzero$ applied to model \eqref{eq:mod2} may be generalized to performing the test on subspaces spanned by the PCs, shown in the \emph{Supporting Information}.  This generalization allows one to perform the association tests on a subset of PCs, while adjusting for other PCs.  It also allows for one to consider rotations of $\bV^T_r$ and projections of $\bV^T_r$ onto relevant subspaces.  For example, it is possible to rotate the PCs to obtain ``independent components'' from independent component analysis (ICA) \cite{ESL2011} and then perform our algorithm on any desired subset of the independent components.  

\section*{Results}   \addcontentsline{toc}{section}{Results}
We evaluated the proposed method on simulated data so that we could directly assess its accuracy, and we also applied the method to two genomic datasets to demonstrate its utility in practice.

\subsection*{Simulation Studies}
Through a set of simulation studies, we demonstrated that the proposed method is able to accurately estimate the statistical significance of associations between the latent variable basis $\bS$ and observed variables $\by_i$ (where $i = 1 \ldots m$). The data in our simulation studies were generated from model \eqref{eq:mod1} $\bY = \bB\bS + \bE$, where variables $\by_i$ corresponding to $\bb_i=\bzero$ are, by definition, the ``null variables'' not associated with $\bS$ (Fig. \ref{LatentVariableModel}). The accuracy of our approach is evaluated by performing $m$ hypothesis tests using the proposed algorithm (where only $\bY$ is observed) and assessing whether the joint distribution of p-values corresponding to the null variables is correctly behaved. 

\subsubsection*{The joint null criterion.} We utilized the ``joint null criterion'' of \cite{LeekStorey2011} to assess whether the set of p-values corresponding to the null variables follow the desired joint distribution (Fig. \ref{SimulationSetup}). When the null hypothesis is true, a valid statistical hypothesis test must generate ``null'' p-values that are distributed uniformly between 0 and 1. We evaluated the uniformity of null p-values using the Kolmogorov-Smirnov test (KS test) which is designed to detect any deviation from the Uniform(0,1) distribution. The set of null p-values produced by a method satisfies the joint null criterion in a given simulation scenario when their joint distribution is equivalent to a set of i.i.d. observations from the Uniform(0,1)  \cite{LeekStorey2011}.  

There are two ways in which we measured deviations from the Uniform(0,1) joint null criterion.  The first is via a two-sided KS test, which detects any deviation; the second is a one-sided KS test, which detects anti-conservative deviations where the null p-values are skewed towards zero.  Anti-conservative deviations will occur when a method does not properly take into account the fact that the association statistics are formed between the variables and PCs (which have been built from the variables themselves), leading to over-fitting and anti-conservative p-values.  Evaluation of the joint null criterion works by simulating many data sets (corresponding to independently repeated studies) from a given data generating process (Fig. \ref{SimulationSetup}).  The joint behavior of the null p-values is then evaluated among these.

We considered $16$ simulation scenarios, described below. For a given scenario, we simulated $500$ independent studies and calculated $500$ KS test p-values, each of which is based on the set of null p-values from its respective study. In other words, for $500$ simulation data sets per scenario, $500$ KS test p-values are calculated to measure deviations from the Uniform(0,1); a second application of the KS test (so-called ``double'' KS test) is then performed on these $500$ KS p-values to assess whether any anti-conservative deviation from the Uniform(0,1) among these studies has occurred (Fig. \ref{SimulationSetup}). If the statistical method being evaluated provides accurate measures of statistical significance, the collection of double KS test p-values must be distributed Uniform(0,1). This guards against any single simulated data set leading one to an incorrect conclusion by chance.    

Overall, we demonstrate that our proposed method provides accurate measures of statistical significance of the associations between variables and the latent variables, when the latent variables themselves are directly estimated from the data via PCA.  At the same time, we show that the conventional method does not provide accurate statistical significance measures.

\subsubsection*{Simulation scenarios and results.}  
\begin{figure*}
\begin{center}
    \includegraphics[width=.8\textwidth]{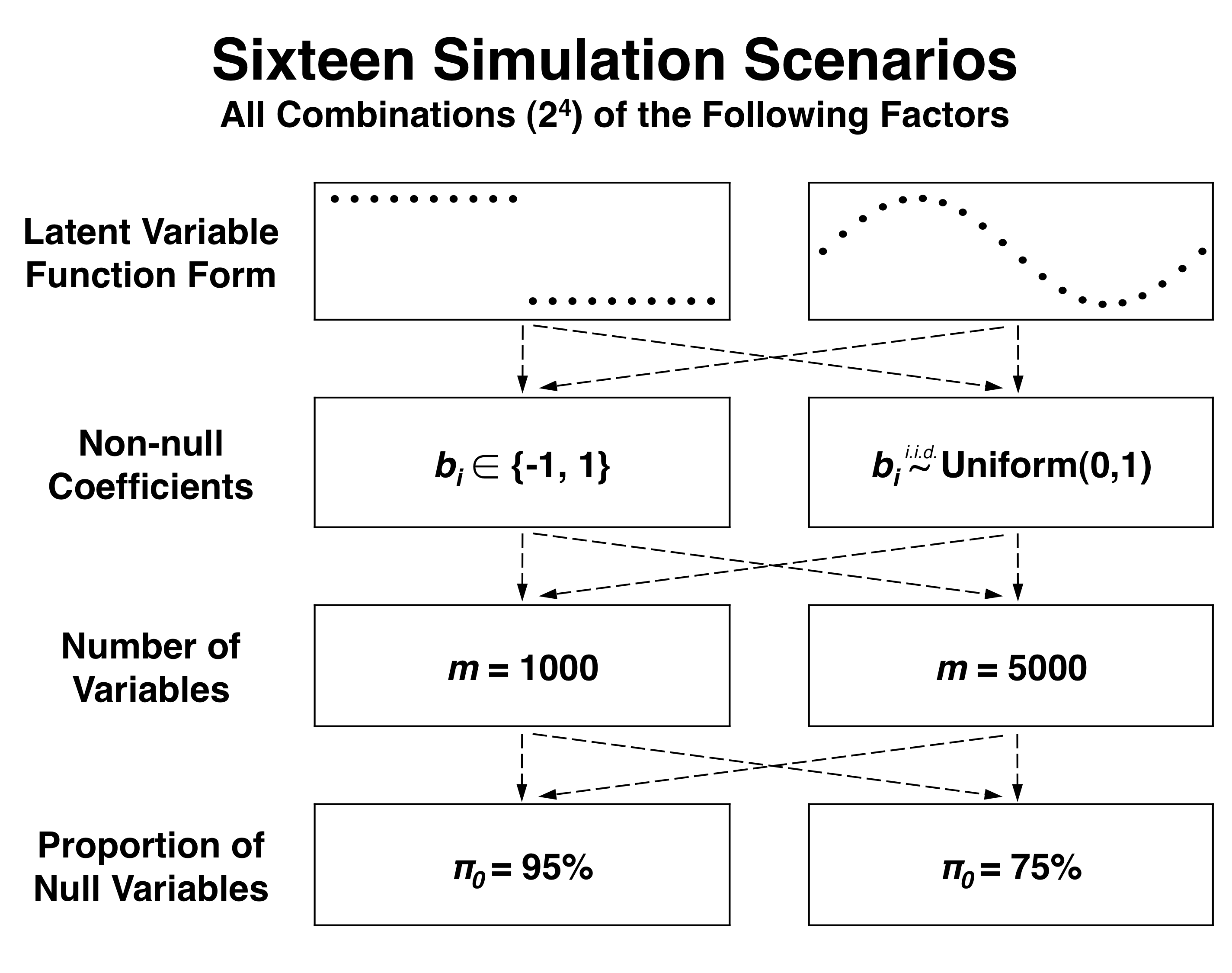}
	\caption{{\em Sixteen simulation scenarios generated by combining four design factors. To assess the statistical accuracy of the conventional F-test and the proposed method, we simulated 500 independent studies for each scenario, and assessed statistical accuracy according to the ``joint null criterion'' \cite{LeekStorey2011}. For a given simulation study, a valid statistical testing procedure must yield a set of null p-values that are jointly distributed Uniform(0,1). We use a KS test to identify deviations from the Uniform(0,1) distribution. Fig. \ref{SimulationSetup} provides a detailed overview of the evaluation pipeline.}}
\label{SimulationScenarios}
\end{center}
\end{figure*}

We constructed $16$ simulation scenarios representing a wide range of configurations of signal and noise (Fig. \ref{SimulationScenarios}), with $500$ independent studies simulated from each. Let's first consider one of the simpler scenarios in detail.  Model \eqref{eq:mod1} is used to generate the data.  In this particular scenario, we have $m=1000$, $n=20$, $r=1$ and
{\small $$ \bS = \frac{1}{\sqrt{n}} (1, 1, 1, 1, 1, 1, 1, 1, 1, 1, \mbox{-}1, \mbox{-}1, \mbox{-}1, \mbox{-}1, \mbox{-}1, \mbox{-}1, \mbox{-}1, \mbox{-}1, \mbox{-}1, \mbox{-}1),$$} 
a dichotomous mean shift resembling differential expression between the first $10$ observations and the second $10$ observations. (The factor $1/\sqrt{n}$ is to give $\bS$ unit variance.) For 95\% of the variables, we set $b_i=0$, implying they are null variables; we parameterize this proportion by $\pi_0=0.95$.  The other 50 non-null variables were simulated such that $b_i \iid$ Uniform(0,1).  The noise terms are simulated as $e_{ij} \iid$ Normal(0,1). The data for variable $i$ are thus simulated according to $\by_i = b_i \bS + \bm{e}_i$.

For a given simulated data set, we tested for the associations between the observed variables and the latent variables by forming association statistics between the observed $\by_1, \by_2, \ldots, \by_m$ and their collective PC, $\bV^T_r$ ($r=1$).  We calculated p-values using both the conventional F-test and the proposed method  with $s=50$ synthetic null variables.  Over $500$ simulated data sets, the conventional F-test resulted in $500$ one-sided KS p-values that exhibit a strong anti-conservative bias with a double KS p-value of $= 9.71 \times 10^{-196}$ (Fig. \ref{SimulationHighlight}, black points). On the other hand, the proposed method correctly calculates null p-values, by accounting for the over-fitted measured error in PCA, with a double KS p-value of $0.502$ (Fig. \ref{SimulationHighlight}, orange points).  Note that the classification of null p-values is based on the true association status from the population-level data generating distribution from model (1), not based on model \eqref{eq:mod2} or on the observed loadings from the PCA.

\begin{figure*}
\begin{center}
    \includegraphics[width=1\textwidth]{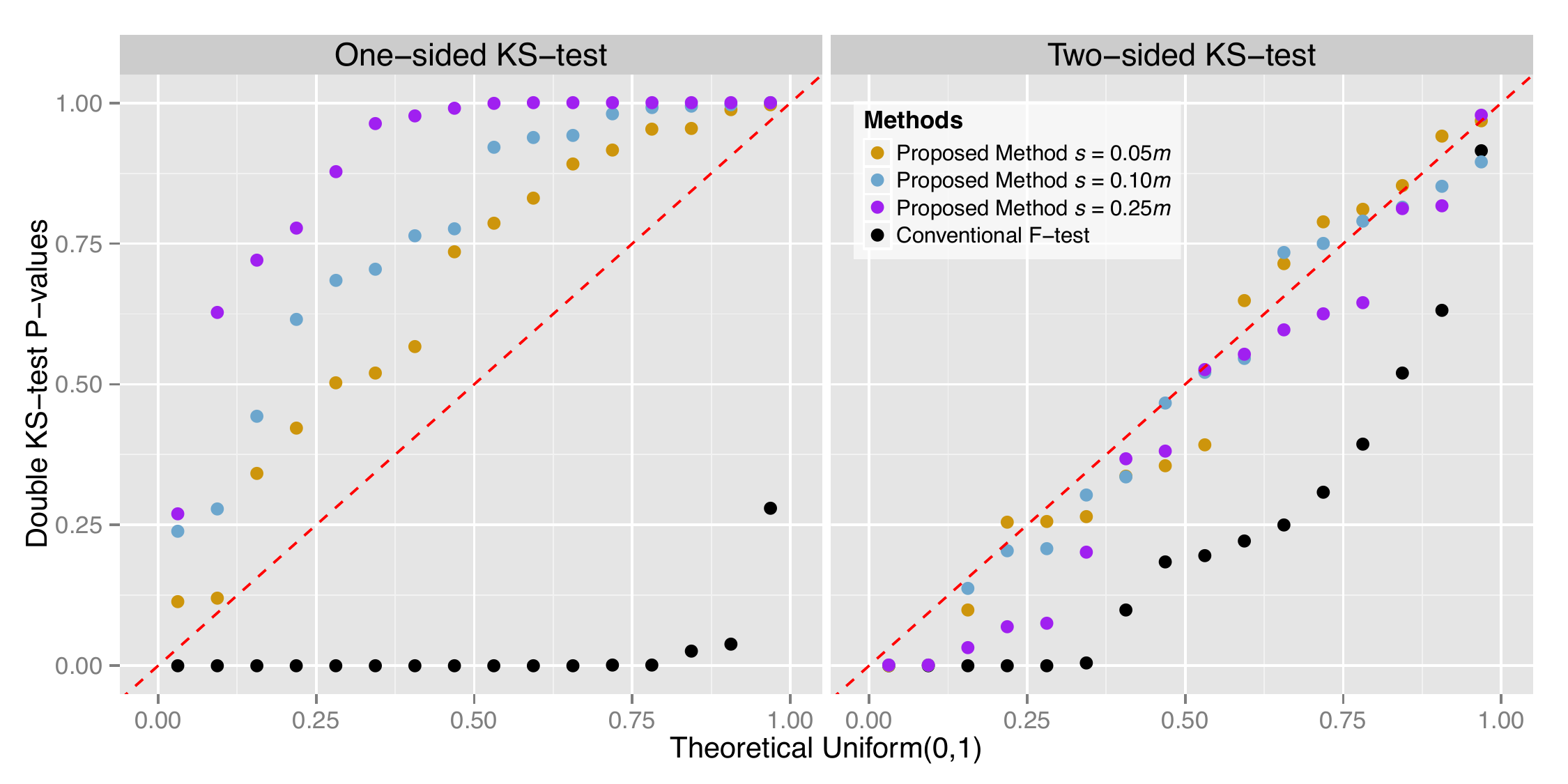}
	\caption{{\em QQ-plots of double KS test p-values from 16 simulation scenarios versus the Uniform(0,1) distribution. For each of 500 independent studies per scenario, we tested for deviation of null p-values from Uniform(0,1), resulting in 500 KS test p-values for each scenario. An individual point in the QQ-plot represents a double KS test p-value for one scenario, comparing its 500 KS test p-values to Uniform(0,1). On the left panel, the systematic downward displacement of 16 black points indicates an anti-conservative bias of the conventional F-test.  In contrast, the proposed method produces null p-values that are not anti-conservative.  On the right panel, a set of 16 points are below the diagonal red line if the joint null distribution deviates from the Uniform(0,1) distribution. The proposed method adjusts for over-fitting of PCA and produces accurate estimates of association significance.}}
\label{QQplots16}
\end{center}
\end{figure*}

We carried out analogous analyses on $15$ more simulation scenarios, detailed in Fig. \ref{SimulationScenarios}. We utilized all possible combinations of the following: (1) either dichotomous or sinusoidal functions for $\bS$; (2) the parameters $\bB$ were simulated from either a Bernoulli or Uniform distribution; (3) $m = 1000$ or $m = 5000$ variables; and (4) the proportion of true null variables set to either $\pi_0 = 0.75$  or $\pi_0 = 0.95$.  The proposed method was applied with $s = 0.05m$, $0.10m$, and $0.25m$ to study the impact of the choice of the number of synthetic null variables. For each scenario, we applied the joint null criterion double KS evaluation (Fig. \ref{SimulationSetup}), using $500$ simulated data sets. The conventional F-test method consistently produced anti-conservative null p-values, while the proposed method yielded accurately distributed null p-values (Fig. \ref{QQplots16}).

In these simulations, we found that the proposed method tended to produce more conservative null p-values as $s$ increased (Fig. \ref{QQplots16}). The explanation for this is that inclusion of a larger number of synthetic null variables leads to a greater over-fitting of PCA to the noise, which in turn yields a conservative empirical null distribution formed by the synthetic null statistics. We therefore identified a trade-off between computational speed and how conservative the calculated p-values are in the choice of $s$.  We note, however, that the null p-values were never observed to be prohibitively conservative in that the power became unreasonably diminished.  In practice, the user has the option to lower the value of $s$ to minimize this, at the cost of greater computation.

We note that we also investigated a delete-$s$ version of the jackstraw, which draws on ideas from our proposed method, which one could call the permute-$s$ jackstraw.  However, this implementation did not produce valid null p-values ({\em Supporting Information}). 

\subsubsection*{Testing for associations on subsets of principal components.}  We have generalized the proposed method to be able to test for associations on any subset of the top $r$ PCs, while adjusting for the remaining PCs among the top $r$.  Here, we demonstrate that the proposed method can identify variables driving a chosen subset of PCs of interest, $\bV^T_{r_1}$, while adjusting for the remaining of the top $r$ PCs which are not of interest, $\bV^T_{r_0}$, where $r_0 + r_1 = r$. Based on model \eqref{eq:mod1}, we simulated data with $m=1000$, $n=20$, $r=2$ and
{\small
\begin{align*}
\bS_1 & = \frac{1}{\sqrt{n}} (1, 1, 1, 1, 1, 1, 1, 1, 1, 1, \mbox{-}1, \mbox{-}1, \mbox{-}1, \mbox{-}1, \mbox{-}1, \mbox{-}1, \mbox{-}1, \mbox{-}1, \mbox{-}1, \mbox{-}1), \\
\bS_2 & = \frac{1}{\sqrt{n}} (1, 1, 1, 1, 1, \mbox{-}1, \mbox{-}1, \mbox{-}1, \mbox{-}1, \mbox{-}1, 1, 1, 1, 1, 1, \mbox{-}1, \mbox{-}1, \mbox{-}1, \mbox{-}1, \mbox{-}1).
\end{align*}
}

$\bS_1$ and $\bS_2$ are truly associated with $100$ variables and $60$ variables, respectively. Among these, $40$ variables that are truly associated with both $\bS_1$ and $\bS_2$. We generated the noise term as $e_{ij} \iid$ Normal(0,1).  We set $r=2$ and tested for associations with the first PC while adjusting for the second PC.  Note that the first PC effectively captured the signal from the first latent variable.  In this case, the null variables were defined to be $900$ variables associated with either only the second latent variable or no latent variable.  The conventional F-test resulted in an anti-conservative bias among the null p-values, with a double KS test p-value of $8.73 \times 10^{-20}$, while the proposed method produced a correct joint null p-value distribution with a double KS test p-value of $0.352$ (Fig. \ref{Simulation_2PC}).

Furthermore, we generalized the proposed method to be able to test for associations on rotations of the top $r$ PCs (and on subsets of these rotations).  This allows for a general exploration of the row-space spanned by the top $r$ PCs.  The details of these generalizations are given in the \emph{Supporting Information}. 

\subsection*{Application to Gene Expression Studies}
Typically, genomic variables are tested for the associations with external variables, which are measured independently of genomic profiling technology, such as disease status, treatment labels, or time points. However, external variables may be imprecise or inaccurate due to poor understanding of the biology or technological limitations; sometimes the external variables of interest may not be capable of being measured at all.  For example, in a cancer gene expression study, the cancer types may be based on histological classification of the tumor cells. Then, association tests, such as F-tests, are conducted between the histological classification and transcriptional levels to discover genes of interest. However, the histological classification of cancer tumors may not distinguish important cancer subtypes \cite{DeRisi1996, Alizadeh2000}.  This lack of information may lead to a spurious signal or reduced power in statistical inference.

When the external variables are unmeasured or imprecise, we are interested in using the latent variable basis, $\bS$, to discover genes of interest (Fig. \ref{MolecularModel}). Because $\bS$ is never directly measured, we must estimate it from the genomic data, using PCA and related methods. We apply our proposed method to two genomic datasets to demonstrate its utility in practice.
 
\subsubsection*{Cell cycle regulated gene expression in Saccharomyces cerevisiae.}
It is known that in \emph{Saccharomyces cerevisiae} there is an abundance of genes whose transcription is regulated with respect to the cell cycle \cite{Spellman1998, Cho1998}. Nonetheless, comprehensive identification of the yeast genes whose expression is regulated by the cell cycle is still an active area of research, since it's unclear how the yeast cell cycle regulation should be quantified and modeled \cite{Tu2005, Pramila2006, Rowicka2007, Wu2008}. The experimental time points after cell population synchronization are readily measured, but this external variable does not directly represent periodic transcriptional regulation with respect to the cell cycle.

Suppose that we want to carry out a hypothesis test on each gene of whether it shows regulation associated with a periodic pattern over the cell cycle. The null hypothesis is then that population mean is not periodic over the cell cycle. This null hypothesis contains an infinite number of mean time-course trajectories that are non-periodic, making the null hypothesis composite.  A composite null hypothesis such as this one is largely intractable because it contains an unwieldy class of potential probability distributions describing gene expression. Indeed, a survey of the literature reveals that this composite null hypothesis is the major challenge when a traditional hypothesis testing approach is taken.  However, by using our approach, we can reduce the complexity of this problem by directly estimating the manifested systematic periodic expression variation and applying the proposed method to identify genes associated with this systematic variation due to the latent variables, $\bS$.

\begin{figure*}
\begin{center}
    \includegraphics[width=.8\textwidth]{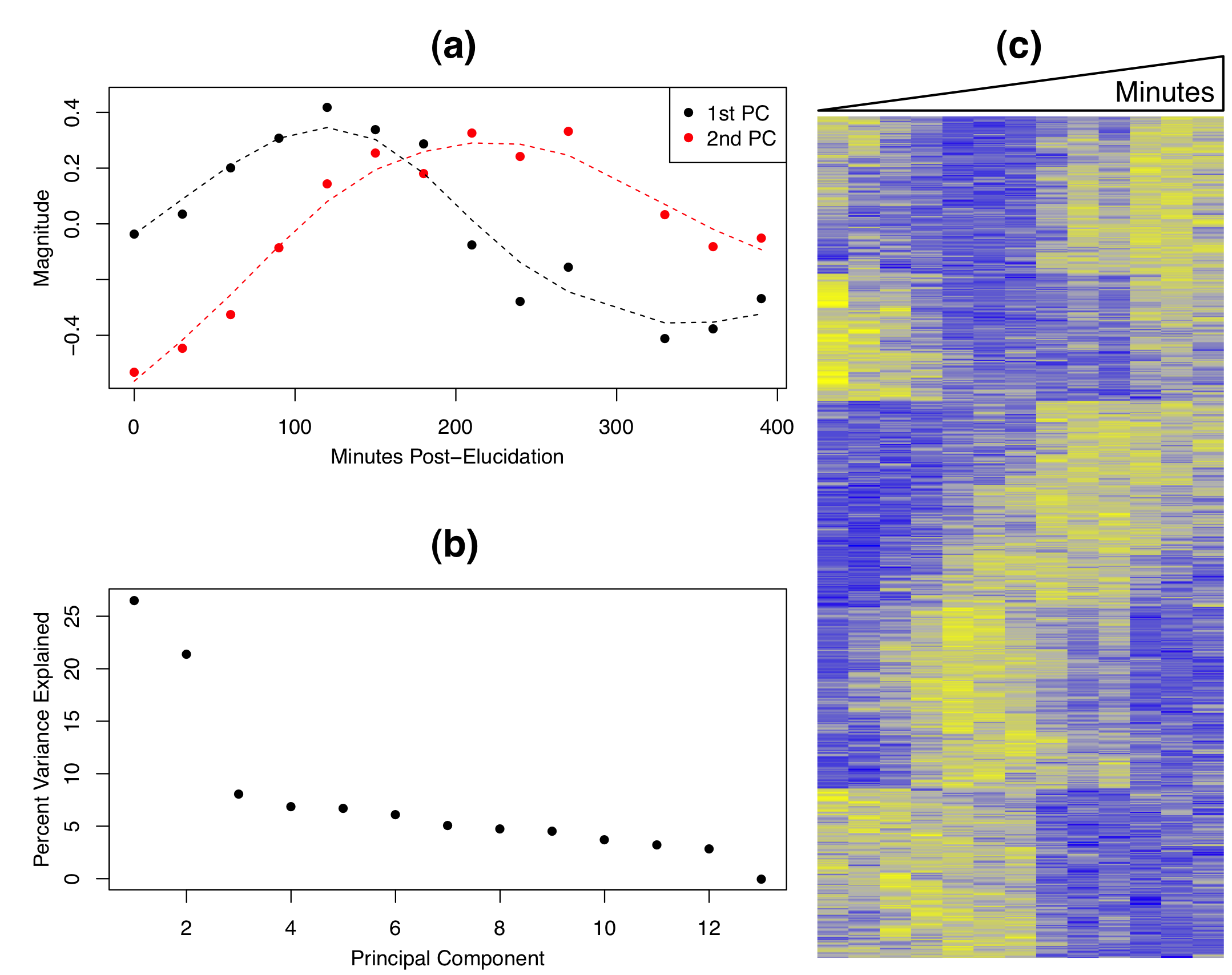}
\caption{{\em Identification of yeast genes associated with the cell cycle regulation. {\rm(a)} The top two PCs of gene expression profiles, measured over time in a population of yeast whose cell cycles have been synchronized by elutriation, capture major transcriptional regulatory patterns \cite{Spellman1998}. {\rm(b)} The percent variance explained by PCs shows that the top two PCs capture 48\% of the total variance in the data. {\rm(c)} Hierarchical clustering of transcriptional levels significantly associated with the top two PCs at $\fdr \leq 1\%$ reveals a diverse set of systematic time-course gene expression trajectories.}}
\label{AlterSVD}
\end{center}
\end{figure*}

\cite{Spellman1998} measured transcriptional levels of $m=5921$ yeast genes, every $30$ min for $390$ min after synchronizing the cell cycle among a population of cells by elutriation\footnote{An inspection of $14$ time-series microarray data from \cite{Spellman1998} reveals an aberrant gene expression profile from $300$ min post-elucidation (Fig. \ref{AlterSVD_OriginalData}). Thus, we decided to remove this array in our re-analysis.}. The top two PCs capture the manifestation of cell cycle regulation on gene expression \cite{Alter2000}, explaining $48\%$ of total variance (Fig. \ref{AlterSVD}a,b). By testing for associations between time-course gene expression and the top two PCs, we essentially transform this challenging problem into a tractable association significance testing problem with a simple null hypothesis $H_0: \bgamma_i = \bzero$ vs. $H_1: \bgamma_i \neq \bzero$ (as opposed to a composite null). The hypothesis test is now simply whether gene $i$ is associated with $r=2$ latent variables estimated by the top two PCs. 

We applied the proposed method to test this hypothesis and identified a large number of genes associated with yeast cell cycle regulation. We discovered that approximately 84.0\% of the $5981$ measured genes are associated with the top two PCs ($\bh{\pi}_0 = 1-0.84 = 0.16$). At $\fdr \leq 1\%$, $2988$ genes were found to be statistically significant. Hierarchical clustering of those $2988$ genes shows strong, yet diverse, periodic patterns (Fig. \ref{AlterSVD}c).  The generalized proposed method allows us to compute statistical significance measures of associations with a subset of PCs. When testing for associations with the first PC while adjusting for the second PC, $1643$ genes were called statistically significant at $\fdr \leq 1\%$, with the estimated proportion of null variables $\widehat\pi_0 = 64.6\%$. On the other hand, at the same $\fdr$ threshold, we found $966$ genes were significantly associated with the second PC with $\widehat\pi_{0} = 59.8\%$.

\subsubsection*{Inflammation associated gene expression in post-trauma patients.}
Large-scale clinical genomic studies often lead to unique analytical challenges, including dealing with a large number of clinical variables, unclear clinical endpoints or disease labels, and expression heterogeneity \cite{leek2007chg}. The ``Inflammation and the Host Response to Injury'' (IHRI) consortium carried out a longitudinal clinical genomics study on blunt force trauma patients.  They collected $393$ clinical variables (some longitudinal) and time-course gene expression (total of $797$ microarrays) on $168$ post-trauma patients \cite{Desai2011}. One of the main goals in this study was to elucidate how inflammatory responses after blunt force trauma are manifested on gene expression. To aggregate relevant clinical variables into a manageable daily score, the IHRI consortium used a modified version of the Marshall score to rate the severity of multiple organ dysfunction syndrome \cite{Marshall1995}.

Based on the modified Marshall score trajectories, \cite{Desai2011} clustered post-trauma patients into five groups, called ``ordered categorical Multiple Organ Failure'' (ocMOF) labels. The time-course gene expression profiles of each patient were summarized by ``within patient expression changes'' (WPEC) \cite{Desai2011}. Then, they tested for correlations between the WPEC genomic variables and the ocMOF score to discover genes associated with inflammatory responses of post-trauma patients. However, the use of the potentially noisy ocMOF clinical variable may impose limitations, as patients with similar Marshall scores may exhibit a wide range of clinical outcomes \cite{Cobb2005}. Furthermore, five discrete values for the ocMOF scores potentially limits the resolution of the clinical variable.

To investigate this further, we utilized our proposed approach where the gene expression itself was used to construct clinical phenotypes on the patients.
We directly used the WPEC data to characterize the molecular signature of inflammatory responses to blunt force trauma. We estimated the manifestation of post-trauma inflammatory responses on gene expression, $\bS$, with the top $9$ PCs (Fig. \ref{ScreeGlue}). Then, we applied the proposed method to identify the genomic variables in WPEC associated with the top $9$ PCs. The original analysis in \cite{Desai2011} estimated $24\%$ of the $54,675$ genomic variables (probe sets) to be associated with the ocMOF score. In contrast, our analysis revealed a much larger proportions of the genomic variables to be significantly associated with the major sources of variation, ranging from $62\%$ for $1^\textnormal{st}$ PC to $39\%$ for $9^\textnormal{th}$ PC.

The genes identified in the original analysis \cite{Desai2011} were largely also identified in our analysis, although our analysis provided many more significant genes. To compare the biological relevance of our re-analysis versus the original analysis, we tested for enrichment of $17$ inflammation-related gene sets \cite{Loza2007}, using one-sided Mann-Whitney-Wilcoxon tests with permutation-based significance. At the $\fdr \leq 1\%$, none of the inflammatory-related gene sets is enriched for the original analysis using the ocMOF scores \cite{Desai2011}. In contrast, a large number of inflammation-related gene sets are significantly enriched when the genomic variables are tested for the associations with the top $9$ PCs individually (Table \ref{Tab:GSA}). MAPK Signaling is enriched for every PC, except $5^\textnormal{th}$ PC, whereas Innate Pathogen Detection is enriched for $1^\textnormal{st}$, $4^\textnormal{th}$, $6^\textnormal{th}$, and $9^\textnormal{th}$ PCs, at the $\fdr \leq 1\%$. Those two biological pathways were emphasized in the original analysis \cite{Desai2011} as indicating down-regulation of innate pathogen detection and up-regulation of MAPK signaling pathway, and they were seen as strong predictors of long-term complications from brute force trauma. Based on enrichment tests, the proposed method appears to provide a biologically richer source of information than the analysis based on the ocMOF scores.   

\section*{Discussion}   \addcontentsline{toc}{section}{Discussion}
We have developed a method to accurately carry out statistical significance tests of associations between high-dimensional variables and estimated latent variables, which have been estimated through systematic variation present in the observed high-dimensional variables themselves. Our approach is to maintain the overall systematic variation in the high-dimensional data set, while replacing a small number of observed variables with independently permuted synthetic null variables. These synthetic null variables allow us to estimate the null distribution of the association statistics calculated on the original data, that takes into account the inherent over-fitting that occurs when estimating latent variables through methods such as PCA. We call this approach the \emph{jackstraw} because it draws on the idea of the game of jackstraws, where a player must remove a stick (i.e., a variable) from a pile of tangled sticks without disturbing the overall structure. Through extensive simulations, we demonstrated that the proposed method is capable of accounting for over-fitting and producing accurate statistical significance measures.  We also demonstrated that applying conventional association testing methods to this problem artificially inflates the statistical significance of associations. 

An input required for the proposed method is the number of PCs, $r$, that capture systematic variation from latent variables. Determining the number of ``statistically significant'' PCs is an active area of research, and defining a number of significant PCs depends on the data structure and the context \cite{Anderson1963, Buja1992, Tracy1996, Johnstone2001, Leek2010}. Setting $r$ too small results in systematic residual variation in model \eqref{eq:mod2}, which hinders the accuracy of the approach. We suggest erring on the side of setting $r$ larger, and utilizing PCs not of interest as adjustment variables in the more general form of our algorithm.  For example, if one would like to identify variables associated with the top three PCs, but is unsure whether the given data has three or four significant PCs, we have found it more robust to input $r=4$ which will adjust for potential systematic residual variation captured by the fourth PC. 

We demonstrated our approach using PCA. It is well known that individual PCs may not be directly interpretable or may contain multiple signals of interest that the user wishes to distinguish.  Our method is capable of considering any subspace spanned by the top $r$ PCs, while adjusting for the remaining subspace spanned by these PCs (to account for systematic variation to of interest).  Our method is also easily adapted to methods such as Independent Component Analysis.  Details are given in the {\em Supporting Information}. This allows the user to intelligently investigate the space spanned by systematic variation and carefully construct estimates of the latent variables of interest.  We do not advocate blindly applying our method to the top $r$ PCs without considering these issues.

The proposed method represents a novel resampling approach operating on variables, whereas established resampling approaches, such as the jackknife and the bootstrap, tend to operate on observations \cite{Quenouille1949, Tukey1958,Efron1979}.  When applying these methods, systematic variation due to latent variables is intentionally perturbed, since their purpose is typically to assess the sampling variation of a single variable. In high-dimensional data, we may need to preserve systematic variation due to latent variables, which is the problem that the jackstraw addresses. 

By accurately testing for associations between observed high-dimensional variables and the systematic manifestation of latent variables in the observed variables, our proposed method allows for the automatic discovery of complex sources of variation and the genomic variables that drive them. The proposed method extends PCA and related methods beyond their popular applications in exploring, visualizing, and characterizing the systematic variation to genomic variable level (e.g., gene-level) significance analyses. Given the increasingly important role that non-parametric estimation of systematic variation plays in the analysis of genomic data \cite{Alter2000, Price2006, leek2007chg}, the proposed method may be useful in many areas of quantitative biology utilizing high-throughput technologies as well as other areas of high-dimensional data analysis.

\subsubsection*{Acknowledgments}
This research was supported in part by NIH grant HG002913 and Office of Naval Research grant N00014-12-1-0764. 

\clearpage
\subsubsection*{Tables} \addcontentsline{toc}{section}{Tables}

\begin{table*}[h]
\caption{Q-values from gene enrichment analysis using inflammation-related gene sets}\label{Tab:GSA}
\scriptsize{\begin{tabular}{ p{4.5cm} p{.7cm} p{.7cm} p{.7cm} p{.7cm} p{.7cm} p{.7cm} p{.7cm} p{.7cm} p{.7cm} | p{.8cm}}
Gene Set & $1^\textnormal{st}$ PC & $2^\textnormal{nd}$ PC & $3^\textnormal{th}$ PC & $4^\textnormal{th}$ PC & $5^\textnormal{th}$ PC & $6^\textnormal{th}$ PC & $7^\textnormal{th}$ PC & $8^\textnormal{th}$ PC & $9^\textnormal{th}$ PC & ocMOF \\ \hline\hline
  \tiny{Adhesion-Extravasation-Migration} & \cellcolor[gray]{0.7}0.004 & 0.034 & 0.053 & \cellcolor[gray]{0.7}0.002 & 0.144 & 0.024 & 0.036 & \cellcolor[gray]{0.7}0.003 & 0.024 & 0.016 \\ 
  \tiny{Apoptosis Signaling} & \cellcolor[gray]{0.7}0.004 & 0.018 & 0.013 & \cellcolor[gray]{0.7}0.004 & 0.036 & \cellcolor[gray]{0.7}0.003 & 0.116 & \cellcolor[gray]{0.7}0.006 & 0.070 & 0.014 \\ 
  \tiny{Calcium Signaling} & 0.021 & \cellcolor[gray]{0.7}0.005 & 0.087 & 0.100 & 0.078 & 0.120 & 0.046 & \cellcolor[gray]{0.7}0.004 & 0.146 & 0.078 \\ 
  \tiny{Complement Cascase} & 0.116 & 0.163 & 0.068 & 0.012 & 0.157 & 0.013 & 0.167 & 0.120 & 0.098 & 0.196 \\ 
  \tiny{Cytokine signaling} & 0.024 & 0.100 & 0.033 & \cellcolor[gray]{0.7}0.007 & 0.140 & \cellcolor[gray]{0.7}0.003 & 0.040 & \cellcolor[gray]{0.7}0.004 & 0.066 & 0.036 \\ 
  \tiny{Eicosanoid Signaling} & 0.020 & 0.031 & 0.042 & \cellcolor[gray]{0.7}0.007 & 0.163 & 0.078 & 0.116 & 0.122 & 0.117 & 0.013 \\ 
  \tiny{Glucocorticoid/PPAR signaling} & 0.100 & 0.034 & 0.040 & 0.027 & 0.182 & 0.039 & 0.041 & \cellcolor[gray]{0.7}0.005 & 0.157 & 0.099 \\ 
  \tiny{G-Protein Coupled Receptor Signaling} & 0.133 & 0.020 & 0.179 & 0.046 & 0.034 & 0.156 & 0.026 & 0.122 & 0.123 & 0.039 \\ 
  \tiny{Innate pathogen detection} & \cellcolor[gray]{0.7}0.004 & 0.077 & 0.018 & \cellcolor[gray]{0.7}0.001 & 0.087 & \cellcolor[gray]{0.7}0.005 & 0.011 & 0.011 & \cellcolor[gray]{0.7}0.007 & 0.039 \\ 
  \tiny{Leukocyte signaling} & \cellcolor[gray]{0.7}0.003 & 0.010 & \cellcolor[gray]{0.7}0.001 & \cellcolor[gray]{0.7}0.002 & 0.044 & \cellcolor[gray]{0.7}0.001 & 0.124 & \cellcolor[gray]{0.7}0.005 & 0.014 & 0.123 \\ 
  \tiny{MAPK signaling} & \cellcolor[gray]{0.7}0.001 & \cellcolor[gray]{0.7}0.002 & \cellcolor[gray]{0.7}0.007 & \cellcolor[gray]{0.7}0.002 & 0.023 & \cellcolor[gray]{0.7}0.002 & \cellcolor[gray]{0.7}0.004 & \cellcolor[gray]{0.7}0.001 & \cellcolor[gray]{0.7}0.002 & 0.036 \\ 
  \tiny{Natural Killer Cell Signaling} & 0.106 & 0.114 & 0.015 & 0.024 & 0.060 & 0.039 & 0.139 & \cellcolor[gray]{0.7}0.004 & 0.036 & 0.167 \\ 
  \tiny{NF-kB signaling} & \cellcolor[gray]{0.7}0.007 & 0.020 & \cellcolor[gray]{0.7}0.007 & 0.017 & 0.120 & \cellcolor[gray]{0.7}0.003 & 0.073 & 0.025 & \cellcolor[gray]{0.7}0.001 & 0.195 \\ 
  \tiny{Phagocytosis-Ag presentation} & 0.025 & 0.064 & 0.010 & 0.011 & 0.098 & 0.020 & 0.013 & \cellcolor[gray]{0.7}0.008 & 0.040 & 0.205 \\ 
  \tiny{PI3K/AKT Signaling} & \cellcolor[gray]{0.7}0.005 & \cellcolor[gray]{0.7}0.001 & 0.071 & 0.059 & 0.163 & 0.011 & \cellcolor[gray]{0.7}0.006 & 0.024 & 0.029 & 0.078 \\ 
  \tiny{ROS/Glutathione/Cytotoxic granules} & 0.016 & \cellcolor[gray]{0.7}0.007 & 0.019 & 0.018 & 0.158 & \cellcolor[gray]{0.7}0.007 & 0.116 & 0.058 & 0.150 & 0.027 \\ 
  \tiny{TNF Superfamily Signaling} & 0.023 & 0.064 & 0.070 & 0.034 & 0.171 & \cellcolor[gray]{0.7}0.007 & 0.159 & \cellcolor[gray]{0.7}0.007 & 0.078 & 0.194\\ \hline\hline
  \multicolumn{11}{p{10cm}}{
  \begin{minipage}{10cm}
    Note: Shaded cells indicate q-value $\leq 0.01$ for a gene set enrichment test.%
  \end{minipage}}
\end{tabular}}
\end{table*}

\clearpage
\subsubsection*{Figures} \addcontentsline{toc}{section}{Figures}

\clearpage
\renewcommand{\thefigure}{S\arabic{figure}}
\renewcommand{\thetable}{S\arabic{table}}

\begin{center}\section*{Supporting Information}\end{center}   \addcontentsline{toc}{section}{Supporting Information}

\subsection*{Generalization of the Proposed Algorithm to Subspaces Spanned by Principal Components.}

{\bf Statistical Hypothesis Tests.} Here, we expand the hypothesis test, $H_0: \bgamma_i = \bzero$ vs. $H_1: \bgamma_i \not= \bzero$, of model \eqref{eq:mod2} to a more general test of a linear hypothesis. Let $\Omega_0$ be the null space of interest and $\Omega_1$ the alternative space.  Suppose that $\bC$ is a $r \times q$ matrix where $r \geq q$ and $\bmm{a}$ is a $q$-vector.  Then $\Omega_0 = \{\bgamma \in \real^r : \bgamma \bC = \bmm{a} \}$ and $\Omega_1 = \{\bgamma \in \real^r : \bgamma \bC \not= \bmm{a} \}$ is a general representation for a null and an alternative space, respectively, of a test of a linear hypothesis.  See \cite{DraperSmith1998} for a thorough treatment of tests of linear hypotheses.

The hypothesis test $H_0: \bgamma_i = \bzero$ vs. $H_1: \bgamma_i \neq \bzero$ for each gene $i$ applied to model \eqref{eq:mod2} can be generalized to
\begin{align*}
H_0 &: \bgamma_i \in \Omega_0 \\
H_1 &: \bgamma_i \in \Omega_1.
\end{align*}
The proposed algorithm can be modified so that the regression based association test performs this hypothesis test (Steps 2 and 5 in the main text algorithm).  This is straightforwardly done by forming F-statistics that are calculated by comparing the unconstrained model fit of model \eqref{eq:mod2} to that under the constraint $\bgamma_i \in \Omega_0$ (see Draper and Smith, 1998).

\begin{figure*}
\begin{center}
    \includegraphics[width=.8\textwidth]{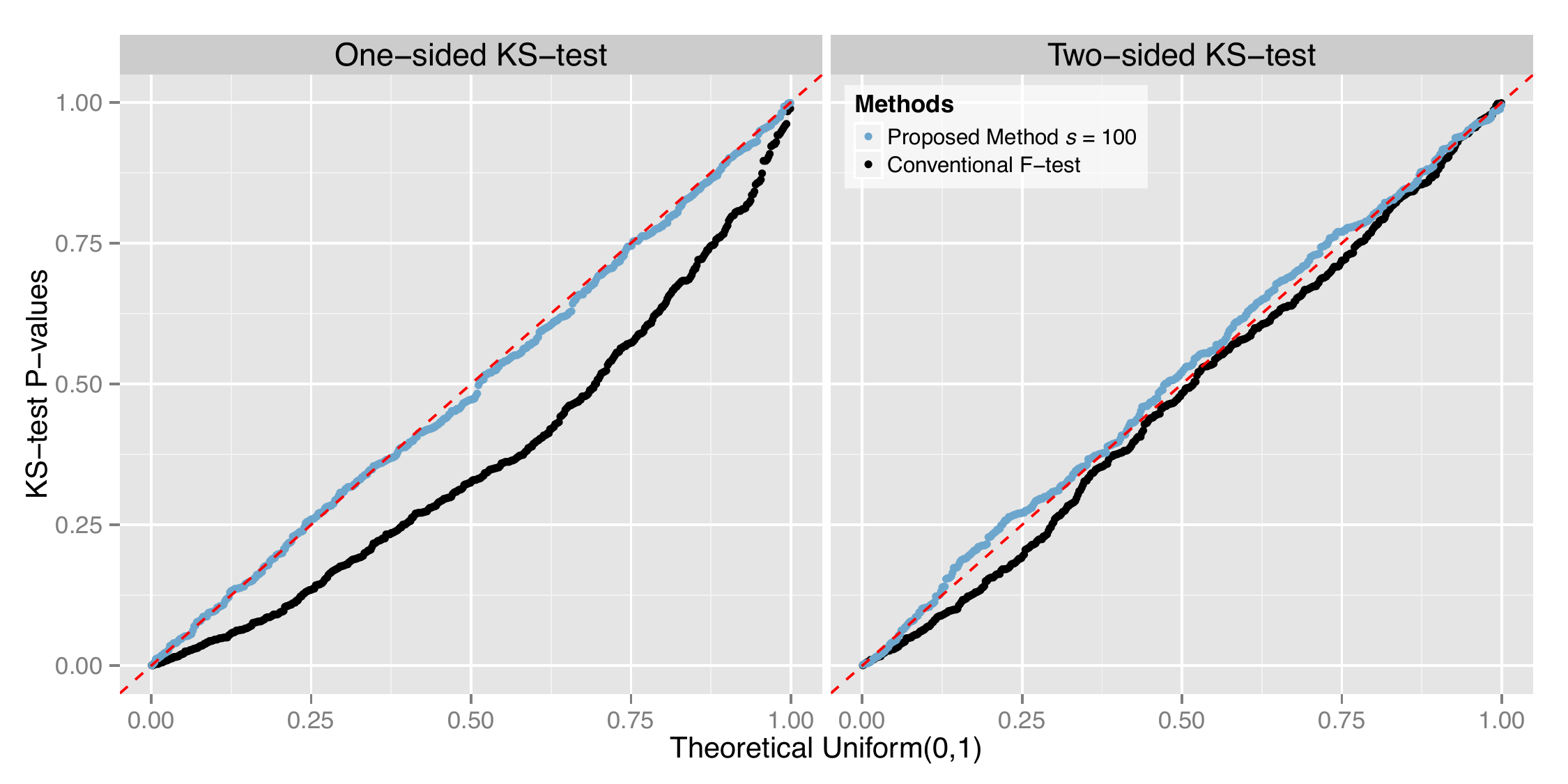}
\caption{{\em Evaluation of statistical tests for the associations on subsets of principal components (PCs). We simulated $m=1000$ variables of $n=20$ observations, with $r=2$ significant PCs. We applied the conventional F-test and the proposed method to test for associations between $m$ variables and the $1^\textnormal{st}$ PC. The proposed method is capable of correctly estimating significance measures of the associations between variables and their $1^\textnormal{st}$ PC, while adjusting for the $2^\textnormal{nd}$ PC. The conventional F-test produces anti-conservative p-values due to over-fitting, where KS test p-values are skewed towards 0.}}
\label{Simulation_2PC}
\end{center}
\end{figure*}

One important procedure enabled by this generalization is to identify variables associated with a subset of rows of $\bV^T_r$. Among multiple statistically significant PCs ($r\geq2$), there may be a subset of PCs of interest, denoted by $\bV^T_{r_1}$. The complementary subset of PCs that are not of interest is denoted by $\bV^T_{r_0}$. The proposed method allows one to estimate the significance of associations between observed variables and $\bV^T_{r_1}$, while adjusting for $\bV^T_{r_0}$. Operationally, $\bV^T_{r_0}$ has to be included in both the constrained and unconstrained models when calculating the F-statistics, whereas $\bV^T_{r_1}$ is only included in the unconstrained model fit. As illustrated in the simulation study, suppose that $r=2$ and we are interested in identifying observed variables associated only with the $1^\textnormal{st}$ PC. In this case, $\Omega_0 = \{(0, \gamma_{i,2}) : \gamma_{i,2} \in \real \}$ and $\Omega_1 = \{(\gamma_{i,1}, \gamma_{i,2}) : \gamma_{i,1} \not=0 \mbox{ and } \gamma_{i,1}, \gamma_{i,2} \in \real \}$. In the proposed algorithm, the $2^\textnormal{nd}$ PC ($=\bV^T_{r_0}$) has to be included in both the constrained and unconstrained models when calculating the F-statistics, whereas the $1^\textnormal{st}$ PC ($=\bV^T_{r_1}$) is only included in the unconstrained model fit.  This scenario with two PCs was simulated, and we demonstrated that the proposed algorithm can produce valid p-values, while accounting for $2^\textnormal{nd}$ PC (right panels in Fig. \ref{Simulation_2PC}). As expected, the conventional F-test results in artificial inflation of statistical significance (left panels in Fig. \ref{Simulation_2PC}).

\  

\noindent {\bf Rotations of Principal Components.} The top $r$ PCs, $\bV^T_r$, collectively capture the systematic variation and estimate the row spanned by $\bS(\bL)$ (Leek, 2010). $\bV^T_r$ may be rotated while spanning the same row space. The proposed method is capable of estimating the statistical significance of associations between observed variables and any rotation of $\bV^T_r$. Consider a $r \times r$ rotation matrix, $\bR$. Then, the generalization of model \eqref{eq:mod2} is
\begin{equation}
\begin{aligned} \label{eq:genmod2}
\bY & = \bGamma \bR\bV^T_r + \bE^{\prime}\\
& = \bGamma \bW_r + \bE^{\prime}.
\end{aligned}
\end{equation}

To perform a significance test of $H_0: \bgamma_i \in \Omega_0$ vs. $H_1: \bgamma_i \in \Omega_1$ for each gene $i$, the algorithm needs to adapt the model \eqref{eq:genmod2} with $\bR\bV^T_r$, in place of model \eqref{eq:mod2} with $\bV^T_r$. The rotation matrix $\bR$ must be applied on $\bV^{*T}_r$ in every iteration of the estimation step of null association statistics. This generalization allows one to perform hypothesis tests using various rotations and projections of PCs.

The rotation matrix $\bR$ could reflect biological or clinical measurements, independent of genomic data. By regressing the top $r$ PCs on a noisy phenotype, we may improve the molecular signature of interest (e.g., separating technical artifacts from biological variation). For example, let's assume a high-dimensional genomic data set contains three statistically significant PCs ($r=3$) and we are interested in identifying variables associated with a particular linear combination of the top three PCs: $\bw_1 = 0.5\bv^T_1 - 0.5\bv^T_2 + \sqrt{0.5}\bv^T_3$. Then, we may construct a $3 \times 3$ rotation matrix $\bR$, which must be orthonormal whose determinant equals 1, as required by any proper rotation matrix. In this case, we arrive at 
\begin{align}
\bR = \begin{bmatrix} 0.5 & -0.5 & \sqrt{0.5} \\
0.5 & -0.5 & -\sqrt{0.5} \\
\sqrt{0.5} &  \sqrt{0.5} & 0 \end{bmatrix}. \nonumber
\end{align}
After obtaining this particular rotation of the top three PCs, $\bW_r = \bR \bV^T_r$, we can continue onto performing a significance test using model \eqref{eq:genmod2}. The two modifications to the algorithm are:

\begin{enumerate}
\item[1.] To apply $\bR$ on both $\bV^T_r$ and $\bV^{*T}_r$ to obtain $\bW_r = \bR \bV^T_r$ and $\bW^*_r = \bR \bV^{*T}_r$ in the computation of observed association statistics and null association statistics, respectively.
\item[2.] To construct a hypothesis test of association between variable $i$ and $\bw_1$, while accounting for $\bw_2$ and $\bw_3$: $\Omega_0 = \{(0, \gamma_{i,2}, \gamma_{i,3}) : \gamma_{i,2}, \gamma_{i,3} \in \real \}$ and $\Omega_1 = \{(\gamma_{i,1}, \gamma_{i,2}, \gamma_{i,3}) : \gamma_{i,1} \not=0 \mbox{ and } \gamma_{i,1}, \gamma_{i,2}, \gamma_{i,3} \in \real \}$.\end{enumerate}

Another interesting example of this generalized jackstraw algorithm arises from transformations of $\bV^T_r$ based on a statistical criterion. Starting with $\bV^T_r$, Independent Component Analysis (ICA) seeks $r$ components that are mutually statistically independent (Comon, 1994). In some cases with statistically independent and non-Gaussian sources of variation, independent components may be more sensible, providing an interpretable low-dimensional representation. We may obtain independent components by rotating $\bV^T_r$ to maximize mutual statistical independence (Hastie et al., 2011).  Let $\bR$ be a rotation matrix that maps $\bV^T_r$ into the $r$ independent components.  To carry out the association test on the $r$ independent components, we use the generalized model \eqref{eq:genmod2} with $\bR\bV^T_r$, instead of model \eqref{eq:mod2} with $\bV^T_r$. As above, the analogous substitution occurs at every iteration to ensure the same rotation is applied to synthetic null variables. 

\

\no \underline{Generalized Algorithm to Calculate Significance of Variables Associated with PCs}
\begin{enumerate}
\item[1.] Obtain $r$ PCs of interest, $\bm{V}^T_r$ by applying SVD to the row-centered matrix $\bY_{m \times n} = \bU\bD\bV^T$.
\item[2.] Rotate $\bV^T_r$ using a $r \times r$ rotation matrix $\bR$ to model $\bY = \bGamma \bR\bV^T_r + \bE^{\prime}$.
\item[3.] Calculate $m$ observed F-statistics $F_1, \ldots ,F_m$, testing $H_0: \bgamma_i \in \Omega_0$ vs. $H_1: \bgamma_i \in \Omega_1$ from the Step 2 model.
\item[4.] Randomly select and permute $s$ rows of $\bY_{m \times n}$, resulting in $\bY^*_{m \times n}$.
\item[5.] Obtain $\bV^{*T}_r$ from SVD applied to $\bY^* = \bU^*\bD^*\bV^{*T}$.
\item[6.] Rotate $\bV^{*T}_r$ using a $r \times r$ rotation matrix $\bR$ to model $\bY^* = \bGamma^* \bR\bV^{*T}_r + \bE^{\prime*}$.
\item[7.] Calculate null F-statistics $F^{0b}_1, \ldots, F^{0b}_s$ from the $s$ synthetic null rows of $\bY^*$, testing $H_0: \bmm{\gamma}^*_i \in \Omega_0$ vs. $H_1: \bmm{\gamma}^*_i \in \Omega_1$ from the Step 6 model.
\item[8.] Repeat steps 4-7 for $b = 1, \ldots, B$ iterations to obtain a total $s \times B$ of null F-statistics.
\item[9.] Compute the p-value for variable $i$ ($i=1,\ldots,m$) by: $$p_i = \frac{\#\{F^{0b}_j \geq F_i; j=1, \ldots, s, b=1, \ldots, B\}}{s \times B}$$
\item[10.] Identify statistically significant tests based on the p-values $p_1, p_2, \ldots,$ $p_m$ (e.g., using false discovery rates).
\end{enumerate}

\  

\noindent {\em Remark 1.} If one wants to test linear hypotheses on the unrotated PCs, then Step 2 is the above algorithm is carried out such that $\bR$ is the identity matrix.

\  

\noindent {\em Remark 2.}  Step 4 of the generalized algorithm above permutes $s$ rows of $\bY$, thereby breaking all systematic variation spanned by $\bV_r^T$ among these $s$ variables.  If one is testing for associations on only a subspace of $\bV_r^T$, then it may be desirable to preserve the systematic variation spanned by the subspace of $\bV_r^T$ acting as adjustment variables.  (E.g., If one is testing for associations with the first PC where $r=2$, then it may be desirable to preserve the systematic variation spanned by the second PC among these $s$ variables.)  This can be accomplished, for example, by modifying the permutations to act only on the residuals after regressing out the adjustment variables, or by carrying out a bootstrap null procedure whereby residuals resulting from this regression are bootstrapped and added back to the adjustment variable only model fit (see, e.g., \cite{ET93}).

\subsection*{An alternative delete-$s$ approach.}  
\begin{figure}[htpb]
\begin{center}
    \includegraphics[width=1\textwidth]{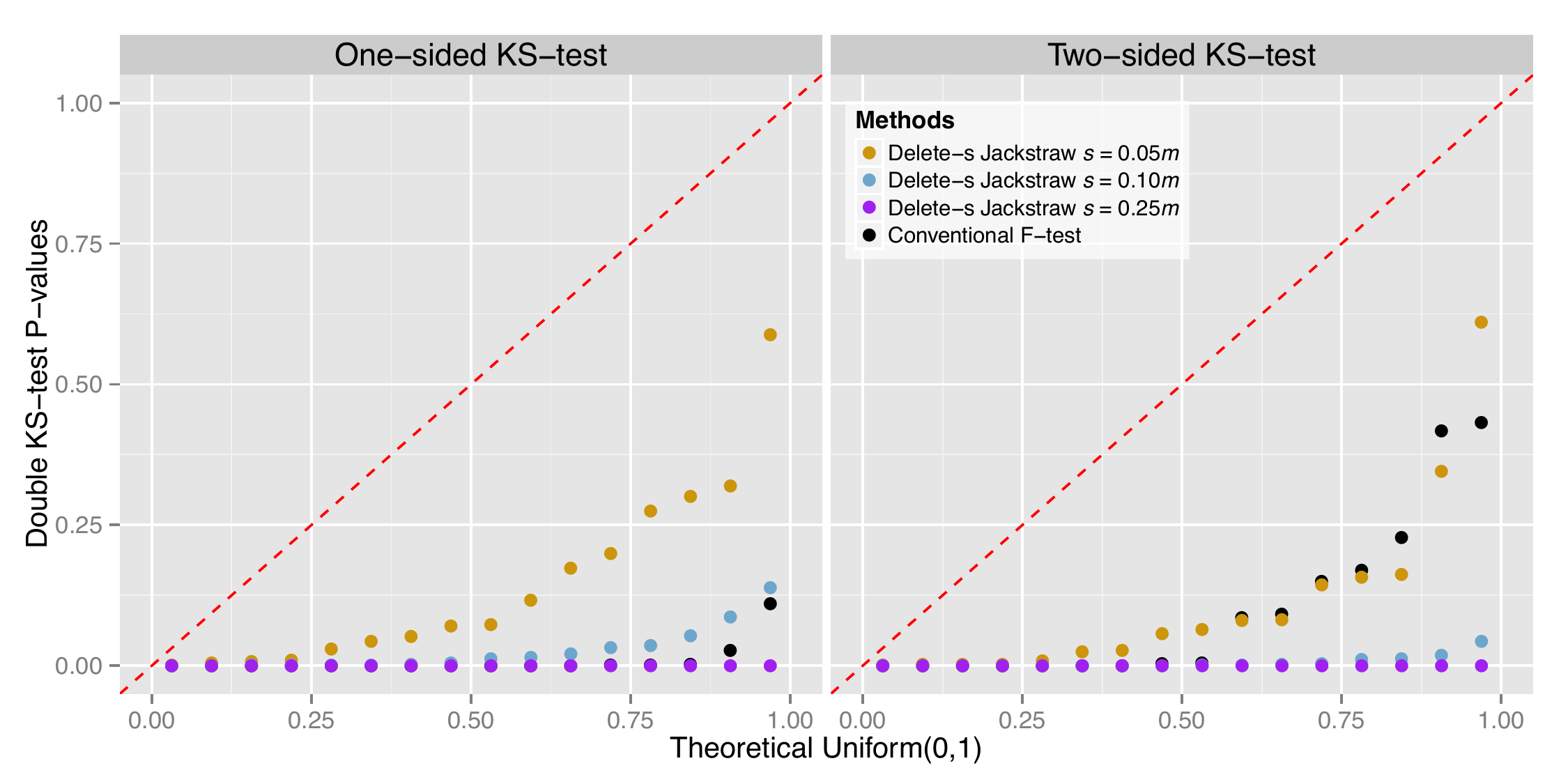}
\caption{{\em QQ-plots of double KS-test p-values from applying the delete-$s$ jackstraw on 16 simulation scenarios. In a delete-$s$ version of the jackstraw, the association p-values between the latent variable basis and a given set of $s$ variables are estimated by testing for the associations between the PCs from $m-s$ variables and the separate set of $s$ variables, which were left out in computation of the PCs. A systematic downward displacement of the points below the red diagonal line indicates anti-conservative p-values, present in both the conventional F-test and the delete-$s$ jackstraw method. A larger value of $s$ in the delete-$s$ jackstraw leads to a greater anti-conservative bias.}}
\label{SimulationStudy_16Scenarios_delete}
\end{center}
\end{figure}

We also investigated a delete-$s$ version of the jackstraw, which draws on ideas from our proposed method, the permute-$s$ jackstraw. In the delete-$s$ jackstraw, the set of $m$ variables is broken up into $m/s$ disjoint sets of $s$ variables each, where $s \ll m$. The significance of associations between a given set of $s$ variables and unobserved latent variables is then calculated by testing for the associations between each of the $s$ variables and the top $r$ PCs calculated among the remaining $m-s$ variables. Since the set of $s$ variables are not used to estimate the latent variables, the association p-values for those $s$ variables should be marginally valid. However, we found that the delete-$s$ jackstraw did not satisfy the joint null criterion. We applied the delete-$s$ jackstraw to the above $16$ simulation scenarios and found that the delete-$s$ jackstraw results in an anti-conservative bias (Fig. \ref{SimulationStudy_16Scenarios_delete}). As $s$ increases, the null p-values from the delete-$s$ jackstraw exhibited a greater anti-conservative bias.  We hypothesis that this delete-$s$ approach did not produce valid null p-values because of a complex dependence among the $m/s$ sets of $s$ p-values (any two sets of $s$ p-values have an overlap of $m-2s$ variables used to construct the PCs used in the association tests), similar to dependence issues encountered in cross-validation.

\clearpage
\subsubsection*{Supplementary Figures} \addcontentsline{toc}{section}{Supplementary Figures}

\begin{figure*}[h]
\begin{center}
    \includegraphics[width=1\textwidth]{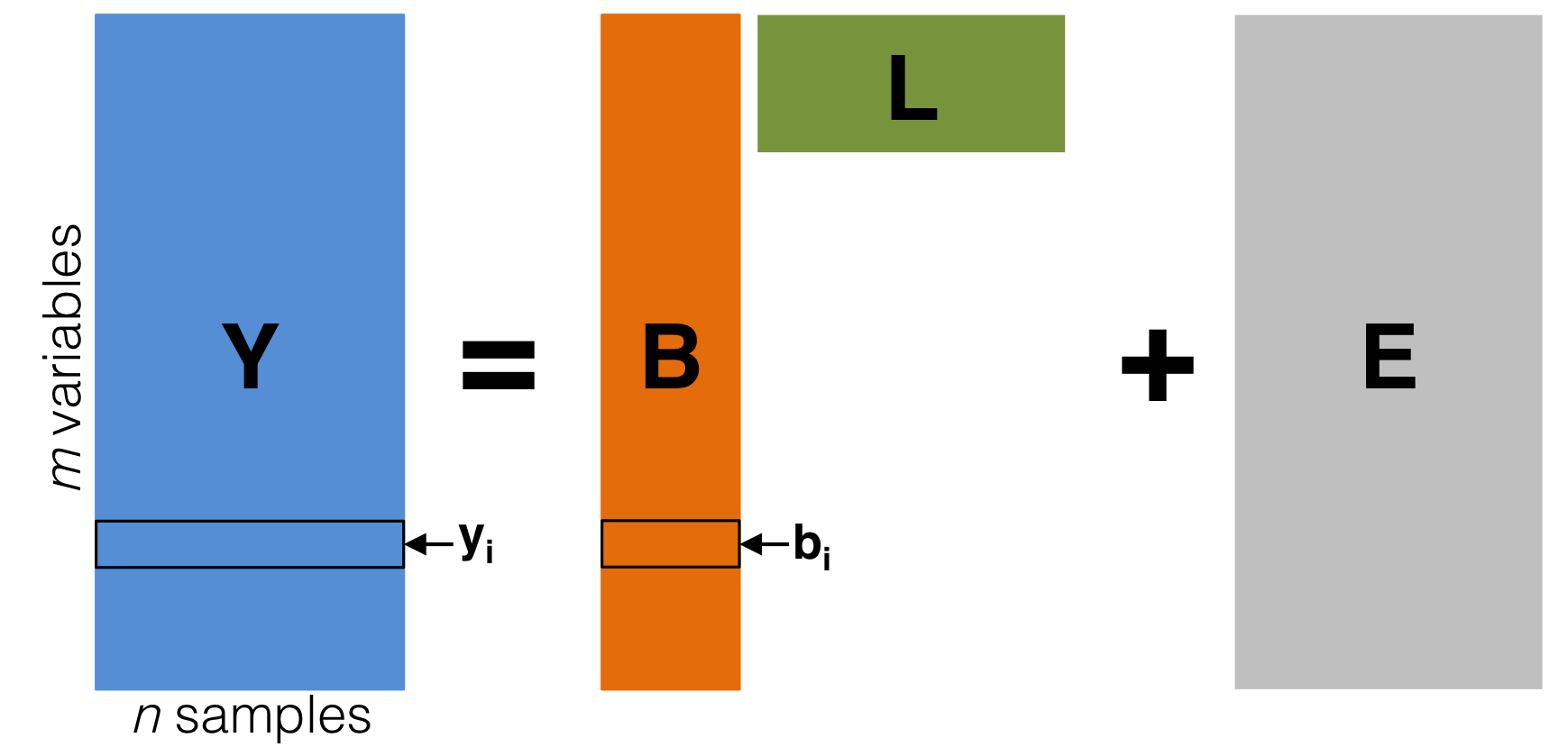}
	\caption{{\em Diagram of the latent variable model \eqref{eq:mod1}. The latent variable basis $\bS$ is not observable, but may be estimated from $\bY$ using the top $r$ principal components ($\bV^T_r$). The noise term $\bE$ is independent random variation. We are interested in performing statistical hypothesis tests on $\bb_i$ ($i=1, \ldots, m$), which quantifies the relationship between $\bS$ and $\by_i$ ($i=1, \ldots, m$). Since $\bS$ must be estimated from $\bY$, a conventional association test results in anti-conservative p-values. We have developed the jackstraw method to account for over-fitting due to using estimates of $\bS$ to compute the statistical significance of associations between $\bS$ and $\by_i$.}}
\label{LatentVariableModel}
\end{center}
\end{figure*}

\begin{figure*}
\begin{center}
    \includegraphics[width=1\textwidth]{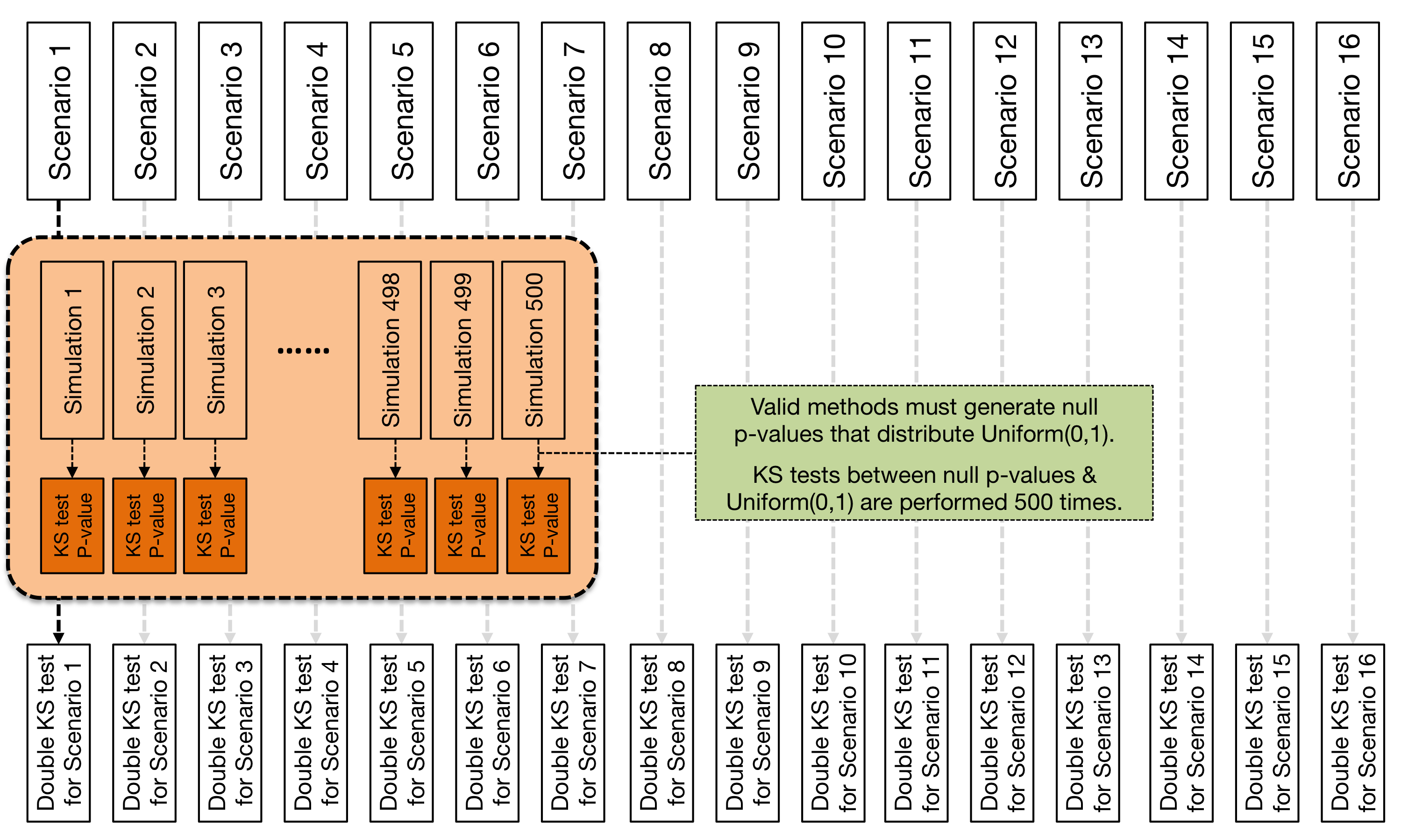}
	\caption{{\em Evaluation pipeline for 16 simulation scenarios. To assess statistical accuracy in testing the associations between variables and principal components, we generated 16 scenarios from a wide range of configurations. For a given scenario, we simulated 500 independent studies, which resulted in 500 KS test p-values (refer to a green box). For a valid statistical method, this set of 500 KS test p-values should be distributed Uniform(0,1); we evaluate an anti-conservative bias among 500 KS test p-values with another application of the KS test (a ``double KS test'') . This results in the 16 simulation scenarios being summarized by 16 double KS test p-values (Fig. \ref{QQplots16}), giving us a comprehensive view of the method's statistical accuracy.}}
\label{SimulationSetup}
\end{center}
\end{figure*}

\begin{figure*}
\begin{center}
    \includegraphics[width=1\textwidth]{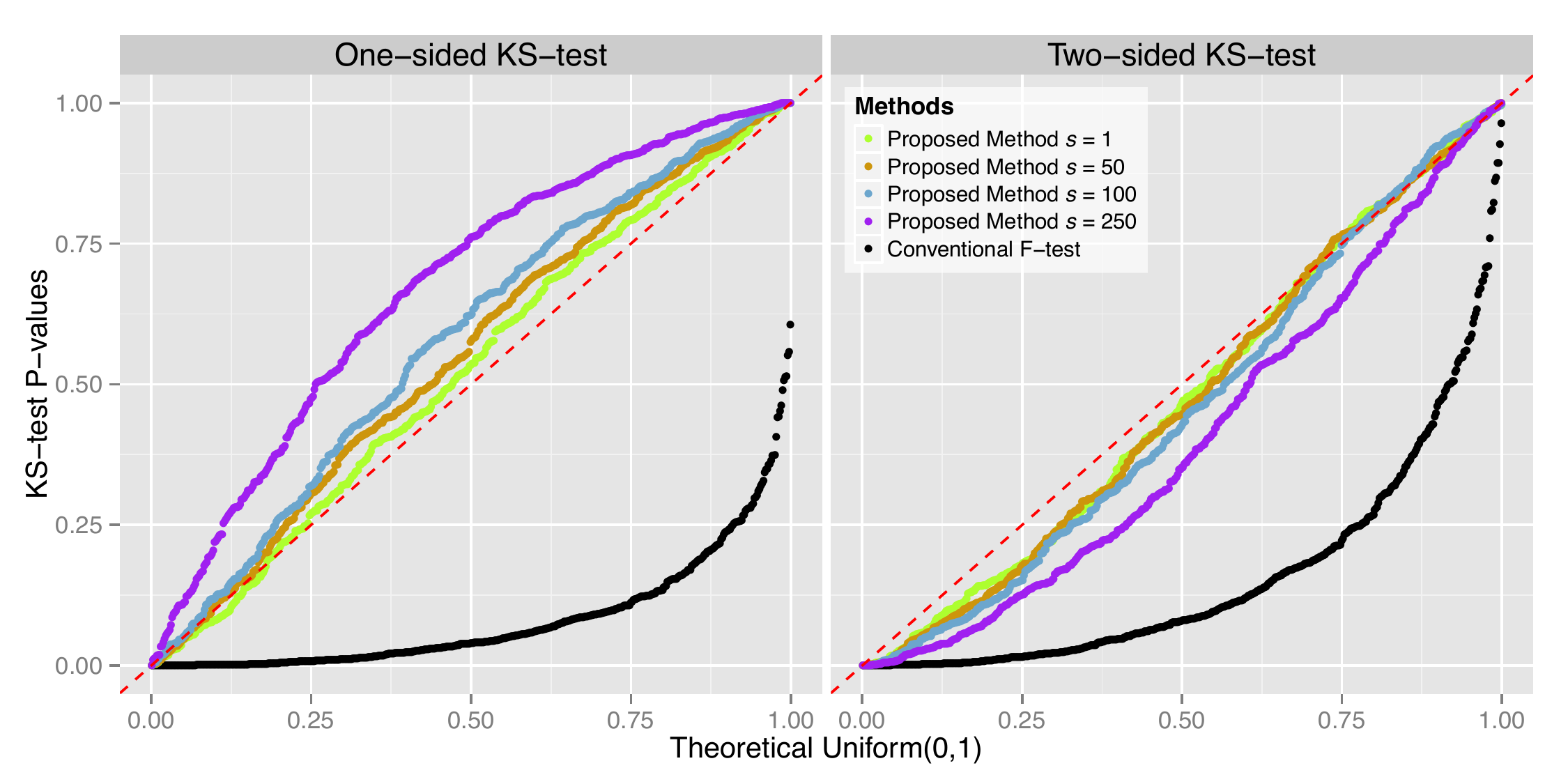}
	\caption{{\em QQ-plots of KS test p-values from the highlighted simulation scenario in the main text, using the conventional F-test and the proposed method. In this scenario, we generated a dichotomous mean shift between two groups of observations with the true proportion of null variables $\pi_0 = 0.95$. We assessed the Uniform(0,1) property of null p-values using the Kolmogorov-Smirnov (KS) test. In the left QQ-plot comparing one-sided KS test p-values and the Uniform(0,1) distribution, the downward displacement of the black points below the diagonal red dashed line indicates anti-conservative p-values resulting from the conventional F-test. In contrast, the upward displacement of the colored points demonstrates how the proposed method guards against anti-conservative bias. On the right panel, a two-sided KS test detects any deviation -- both conservative and anti-conservative -- from the theoretically correct Uniform(0,1) distribution.}}
\label{SimulationHighlight}
\end{center}
\end{figure*}

\begin{figure}[htpb]
\begin{center}
    \includegraphics[width=.5\textwidth]{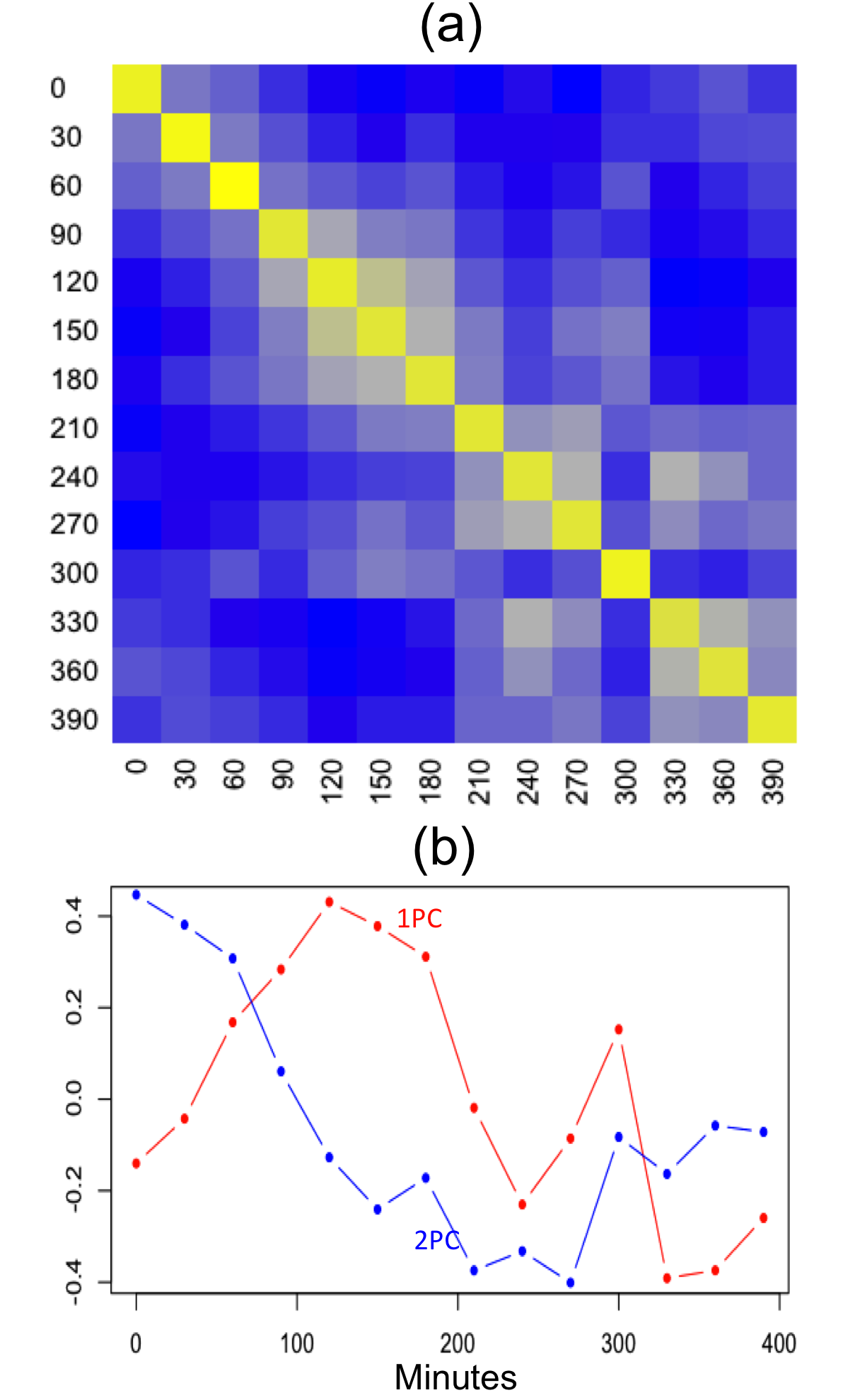}
\caption{{\em Original gene expression profiles of the Spellman \emph{et al.} (1998) yeast cell cycle experiment. {\rm(a)} A heat map of the covariance matrix of 14 arrays reveals an outlier from the time point at 300 min (low values in blue and high values in yellow). {\rm(b)} The top two PCs of the original 14 microarrays corroborate an aberrant gene expression profile from 300 min, which is removed from our new analysis.}}
\label{AlterSVD_OriginalData}
\end{center}
\end{figure}

\begin{figure}[htpb]
\begin{center}
    \includegraphics[width=.8\textwidth]{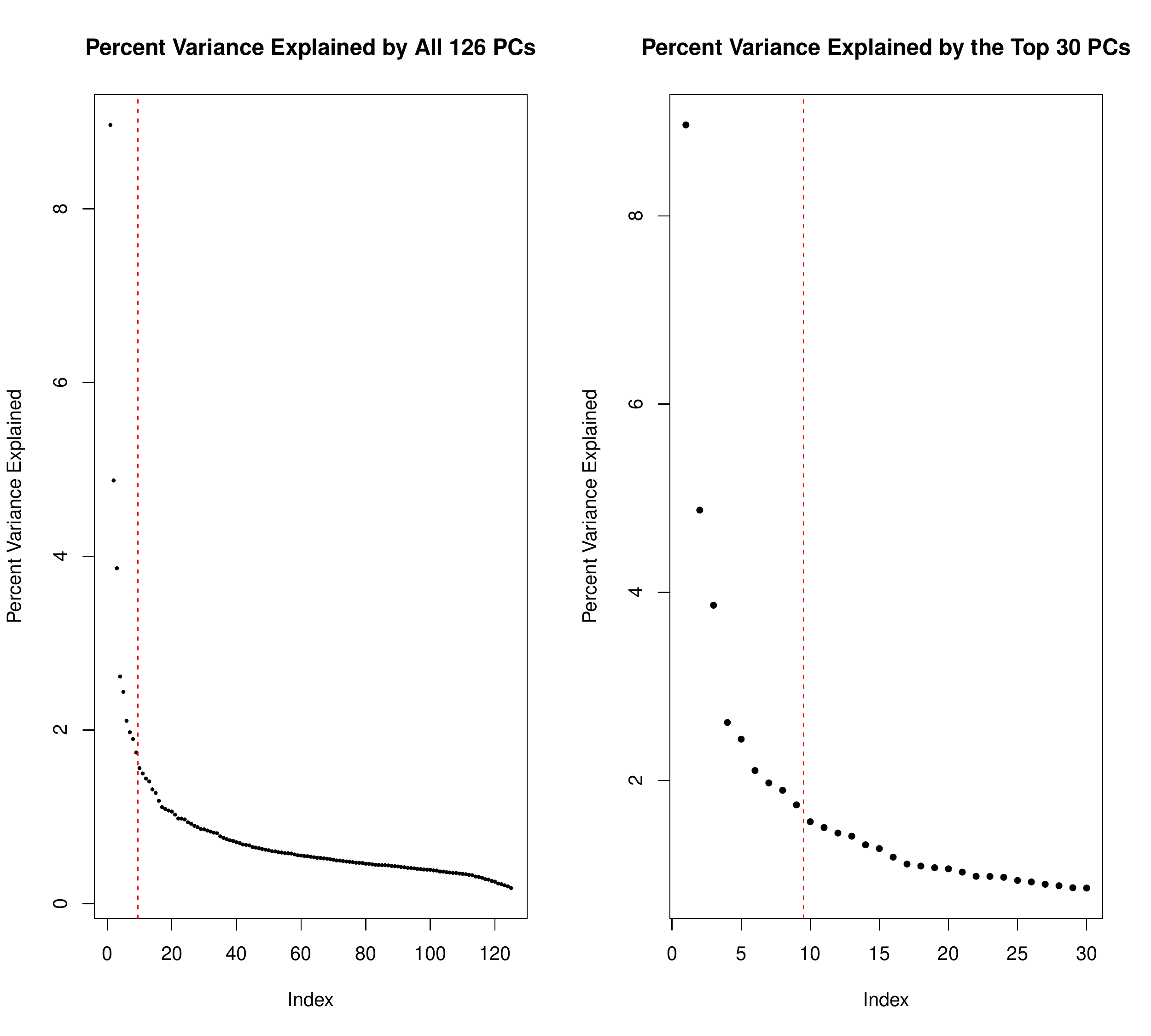}
\caption{{\em Percent variance explained by principal components (PCs) of the ``Within Patient Expression Change (WPEC)'' matrix from Desai \emph{et al.} (2011). The considerable drop (the ``elbow'') in this scree plot helped us to identify the top 9 PCs as significant, likely capturing the molecular signatures of patient responses to blunt force trauma.}}
\label{ScreeGlue}
\end{center}
\end{figure}

\clearpage
\phantomsection
\addcontentsline{toc}{section}{References}
\bibliography{refs}
\bibliographystyle{nature}

\end{document}